\documentclass[12pt]{iopart}

\usepackage{iopams}  
\usepackage{amsmath}
\usepackage{amsthm}
\usepackage{amssymb}
\usepackage{bbold}
\usepackage[toc,page]{appendix}
\usepackage{graphicx}

\begin{document}

\title{Calculation of time resolution of the J-PET tomograph using the Kernel Density Estimation}

\author{L~Raczy\'nski$^1$, W~Wi\'slicki$^1$, W~Krzemie\'n$^2$, P~Kowalski$^1$, D~Alfs$^3$, T~Bednarski$^3$,
P~Bia\l{}as$^3$, C~Curceanu$^4$ E~Czerwi\'nski$^3$, K~Dulski$^3$, A~Gajos$^3$, B~G\l{}owacz$^3$, M~Gorgol$^5$, B~Hiesmayr$^6$, B~Jasi\'nska$^5$, D~Kami\'nska$^3$, G~Korcyl$^3$, T~Kozik$^3$, N~Krawczyk$^3$, E~Kubicz$^3$, M~Mohammed$^3$, 
M~Pawlik-Nied\'zwiecka$^3$, S~Nied\'zwiecki$^3$, M~Pa\l{}ka$^3$, Z~Rudy$^3$, O~Rundel$^3$, N~G~Sharma$^3$, M~Silarski$^3$, J~Smyrski$^3$, A~Strzelecki$^3$, A~Wieczorek$^3$, B~Zgardzi\'nska$^5$, M~Zieli\'nski$^3$, P~Moskal$^3$}

\address{$^1$ Department of Complex System, National Centre for Nuclear Research, 05-400 Otwock-\'Swierk, Poland}
\address{$^2$ High Energy Physics Division, National Centre for Nuclear Research, 05-400 Otwock-\'Swierk, Poland}
\address{$^3$ Faculty of Physics, Astronomy and Applied Computer Science, Jagiellonian University, 30-348 Cracow, Poland}
\address{$^4$ INFN, Laboratori Nazionali di Frascati, 00044 Frascati, Italy}
\address{$^5$ Institute of Physics, Maria Curie-Sk\l{}odowska University, 20-031 Lublin, Poland}
\address{$^6$ Faculty of Physics, University of Vienna, 1090 Vienna, Austria}

\ead{lech.raczynski@ncbj.gov.pl}
\vspace{10pt}
\begin{indented}
\item[]07 April 2017
\end{indented}

\begin{abstract}
In this paper we estimate the time resolution of the J-PET scanner built from plastic scintillators. We incorporate the method of signal processing using the Tikhonov regularization framework and the Kernel Density Estimation method. We obtain simple, closed-form analytical formulas for time resolutions. The proposed method is validated using signals registered by means of the single detection unit of the J-PET tomograph built out from 30 cm long plastic scintillator strip. It is shown that the experimental and theoretical results, obtained for the J-PET scanner equipped with vacuum tube photomultipliers, are consistent. 

\end{abstract}

%
\vspace{2pc}
\noindent{\it Keywords}: Positron Emission Tomography, Time Resolution, Kernel Density Estimation
%
%
%
%

\section{Introduction}

\begin{figure}[h!]
	\centerline{\includegraphics[width=0.45\textwidth]{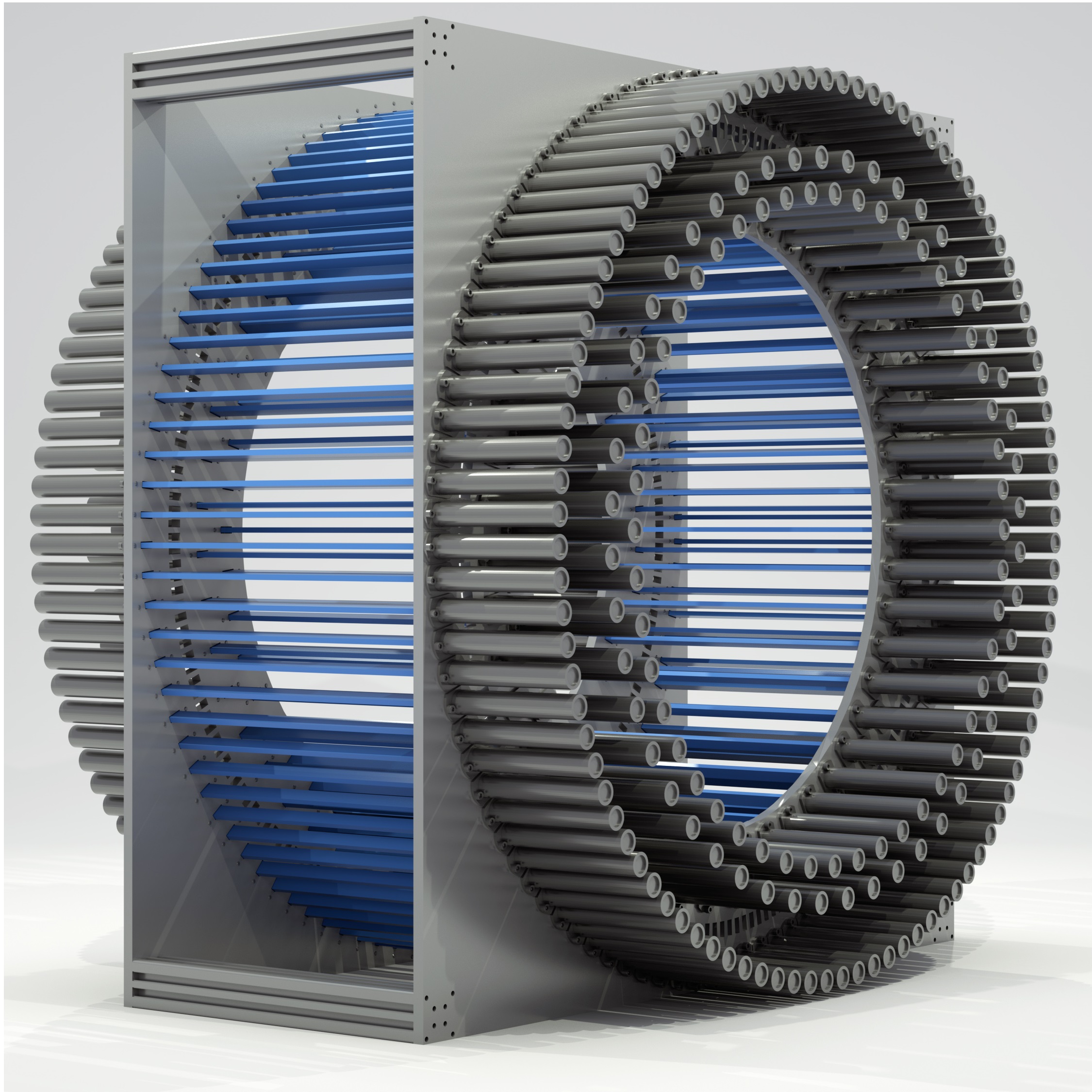}}
\caption{Schematic visualization of an example of three layer J-PET detector. Each scintillator strip is aligned axially and read out at two ends by photomultipliers. 
\label{Fig:JPETbarrel}
}
\end{figure}

The Jagiellonian PET (J-PET) Collaboration constructs a PET scanner from plastic strips forming the barrel~(Moskal \etal 2011, 2014a). 
An example of arrangement of the scintillator strips in the J-PET tomograph is visualized in Fig.~\ref{Fig:JPETbarrel}. The proposed setup permits to use more than one detection layer thus increasing the efficiency of $\gamma$ photon registration~(Moskal \etal 2016). A single detection module consists of a long scintillator strip and a pair of photomultipliers attached to the opposite ends of the strip. Measurement with such detector results in timestamps from both sides of each scintillator, which allow to extract the timing, position and energy information of each $\gamma$ photon interaction. The time and position of the $\gamma$ photon interaction in the scintillator strip is calculated based on times at left ($t_{(L)}$) and right ($t_{(R)}$) side of the strip. In the first approximation, the time of interaction may be estimated as an arithmetic mean of $t_{(L)}$ and $t_{(R)}$ and the position of interaction along the strip may be calculated as $(t_{(L)} - t_{(R)}) v/2 $, where $v$ denotes the speed of light signals in the scintillator strip. The energy deposited in the scintillator strip may be expressed in terms of the number of photoelectrons registered by the photomultipliers and is proportional to the arithmetic mean of a number of photoelectrons registered at the left and right sides of the scintillator; the value of  energy calibration factor was evaluated in Ref.~(Moskal \etal 2014b). The registration of single event of positron emission, used for the image reconstruction, is based on the detection of both $\gamma$ photons in two modules in a narrow time window. Therefore, a single image-building event includes information about four times of light signals arrival to the left and right ends of the two modules that register in coincidence.
The J-PET detector offers the Time of Flight (TOF) resolution  competitive to existing solutions (Humm \etal 2003, Townsend \etal 2004, Karp~\etal 2008, Conti~2009, Conti~2011, 
S\l{}omka~\etal 2016), due to the fast plastic scintillators and dedicated electronics allowing for sampling in the voltage domain of signals with durations of few nanoseconds (Pa\l{}ka \etal 2014). 

Recently, the time resolution, defined hereafter as the standard deviation, of about 80~ps has been achieved for the registration of $\gamma$ photon in 30~cm long scintillator strips read out at both ends by the vacuum tube photomultipliers (Moskal \etal 2014b, Raczynski \etal 2014). Such resolution results in coincidence resolving time (CRT) of about 275~ps as  shown in Ref.~(Moskal \etal 2016). Further improvement of time resolution requires developments in techniques of signal processing and effective parametrizations of detector's features. Our estimate of the time resolution is based on statistical properties of the signals in plastic scintillators. Distribution of the time of the photon emission followed by its interaction in plastic scintillators was described in Refs.~(Moszynski, Bengtson 1977, 1979). Following the time order statistics analysis described e.g. in Refs.~(Seifert \etal 2012, Degrot 1986, Spanoudaki, Levin 2011), the statistical framework allowing for the analysis of photon propagation in the scintillator strips was proposed in Ref.~(Moskal \etal 2016). 

In this paper we propose a novel approach to calculate the time resolution of the PET scanner based on ideas from the Tikhonov regularization (Tikhonov 1963, 1977) and Kernel Density Estimation (Parzen 1962, Rosenblat 1956) methods. We investigate the quality of estimation of time resolution based on the scheme with a single scintillator strip detector introduced in Refs.~(Moskal \etal 2014b, Raczynski \etal 2014). The most important aspect of the time resolution evaluation involves the statistical description of noise. The noise in the measured signal comprises two components: statistical fluctuations of the number of photoelectrons registered by the photosensor, and effect of the limited number of samples of the signal in the voltage domain. In Ref.~(Raczynski \etal 2015a), the formula for calculations of the signal recovery error was introduced and proven. In this paper we determine dependence of the signal estimation error on the number and shape of registered photoelectron signals. Theoretical results are compared to the experimental resolutions achievable using traditional readout with the vacuum tube photomultipliers. The method is verified by setting in calculations the same conditions as in the experiment, as described in Refs.~(Moskal \etal 2014b, Raczynski \etal 2014). 

The J-PET tomograph can be equipped with various types of photomultipliers: the vacuum tube photomultipliers (standard in the 
J-PET prototype), the silicon or the microchannel plates photomultipliers. In case of the vacuum tube and silicon photomultipliers, the registration of the whole signal is not possible, and therefore sampling in the voltage domain using a predefined number of voltage levels is needed. The output signal is then recovered using ideas from the Tikhonov regularization (Tikhonov 1963, 1977) and compressive sensing (Candes \etal 2006, Donoho 2006) methods. The microchannel plates photomultipliers are the most promising in view of the application in the J-PET instrument due to the possibility of direct registration of a timestamp of each single photon. In the experimental study we will derive time resolutions of various configurations of the J-PET detector using different types of photomultipliers.
  
\section{Materials and methods} \label{Sec:MatMeth}

In this work, we assume that the $\gamma$ photon interacts in the scintillator strip at time $\Theta$ and in the position $x$. We consider resolution for these reconstructions. 

The time of the photon registration at the photomultiplier, referred to as $t_r,$ is considered as a random variable, equal to the sum of three contributing values:
\begin{equation}
	t_r = t_e + t_p + t_d, 	\label{Eq:t_r}
\end{equation}
where $t_e$ is the photon emission time, $t_p$ is the propagation time of the photon along the scintillator strip and $t_d$ is the photomultiplier transit time. Assuming that the times $t_e, t_p, t_d,$ given in Eq.~(\ref{Eq:t_r}), are independent random variables with probability density functions (pdfs) denoted with $f_{t_e}, f_{t_p}, f_{t_d},$ respectively, the distribution function of $t_r$ is given as the convolution: 
\begin{equation}
	f_{t_r}(t) = (f_{t_e} * f_{t_p} * f_{t_d})(t), \quad t > 0. \nonumber
\end{equation}

In case of the ternary plastic scintillators used in the J-PET detector (Saint Gobain Crystals, Eljen Technology), the distribution of $t_e$ is well approximated by the following formula (Moszynski, Bergston 1977, 1979):
\begin{equation}
	f_{t_e}(t) = \kappa_e \int_{\Theta}^{t} \left( e^{-\frac{t-\tau}{\tau_d}} - e^{-\frac{t-\tau}{\tau_r}} \right)
	e^{-\frac{(\tau-\Theta -2.5\sigma_{e})^2}{2\sigma^2_e}} d\tau,
	\label{Eq:f_t_e}
\end{equation}
where $\tau_d = 1.5$~ns, $\tau_r = 0.005$~ns and $\sigma_e = 0.2$~ns, and $\kappa_e$ stands for the normalization constant. The values of the parameters $\tau_d, \tau_r, \sigma_e$ were adjusted in order to describe the properties of the light pulses from the BC-420 scintillator (Moskal \etal 2016, Saint Gobain Crystals). By definition in Eq.~(\ref{Eq:f_t_e}):
\begin{equation}
	t_e \geqslant \Theta.	\label{Eq:condition_f_t_e}
\end{equation}

\begin{figure}[h!]
	\centerline{\includegraphics[width=0.6\textwidth]{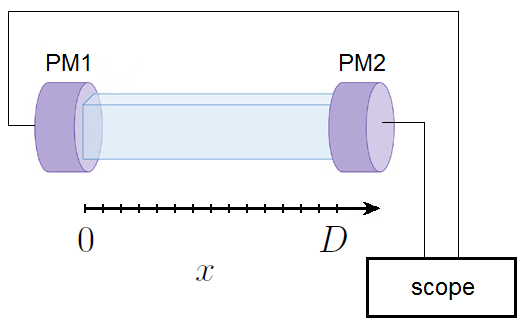}}
\caption{Measurement provided with single scintillator strip. The variable $x$ describes the position of the emission point along the strip. 
\label{Fig:exp_setup}
}
\end{figure}

Initial direction of flight of the photon in the scintillator is uniformly distributed. The photon on its way along the scintillator strip from the emission point to the photomultiplier may undergo many internal reflections whose number depends on the scintillator's geometry and the photon's emission angle. However, the space reflection symmetries of the cuboidal shapes, considered in this article, enables a significant simplification of the photon transport algorithm, without following photon propagation in a typical manner. The statistical modelling of this phenomena was presented in details in Ref.~(Moskal \etal 2016) and the analytical function describing the distribution function $f_{t_p}$ may be expressed by the following formula:
\begin{equation}
	f_{t_p}(t) = \frac{\kappa_p \cdot x}{t^2} \cdot e^{-\mu_{\text{eff}} \cdot v \cdot t},
	\label{Eq:f_t_p}
\end{equation}
where $v$ is the speed of light in the scintillator strip, $\mu_{\text{eff}}$ is the effective absorption coefficient for the scintillator material and $\kappa_p$ the normalization constant. The $0 \leqslant x \leqslant D$ is the longitudinal position of the emission point (see Fig.~\ref{Fig:exp_setup}). The pdf function $f_{t_p}(t)$ in Eq.~(\ref{Eq:f_t_p}) is nonzero only for:
\begin{equation}
	t_p \geqslant \frac{x}{v},
	\label{Eq:condition_f_t_p}
\end{equation}
where $t_p = \frac{x}{v}$ corresponds to the photon flying along the strip.

Finally, the time of registration $t_r$ is smeared using Gaussian distribution centered on the mean transition time $T_d$ and variance $\sigma_d^2$ estimated empirically:
\begin{equation}
	f_{t_d}(t) = \frac{1}{\sqrt{2 \pi }\sigma_d } \exp \left( - \frac {(t - T_d)^2}{\sigma_d^2} \right).
	\label{Eq:f_t_d}
\end{equation}

In this work, we assume that the signal registered at the photomultiplier output has the same functional dependence on the time as the $f_{t_r}$ function. We assume that the signal $y \in R^N$ is discretized by the oscilloscope. It is sampled in the constant time intervals denoted with $T_s.$ From the conditions Eq.~(\ref{Eq:condition_f_t_e}) and (\ref{Eq:condition_f_t_p}), it follows that the registration time $t_r$ fulfils the inequality:
\begin{equation}
t_r \geqslant \Theta + \frac{x}{v}.		\nonumber
\end{equation}
It was assumed that the transition time $t_d \geqslant 0.$ Therefore, the $n^{th}$ time sample is given by:
\begin{equation}
	t^{(n)} = nT_s + \Theta + \frac{x}{v} \quad \quad n = 1, 2, ..., N,				
	\label{Eq:t_n_def}
\end{equation}
and the $n^{th}$ sample of the signal $y$ is given as:
\begin{equation}
	y(n) = \beta(E, x) \cdot f_n, \quad \text{where} \quad f_n = f_{t_r}(t^{(n)}) \quad \quad n = 1, 2, ..., N,		\label{Eq:def_y}									
\end{equation} 
where $\beta(E,x)$ is a coefficient providing the scaling of the pdf function $f_{t_r}$ in order to obtain the voltage signal:
\begin{equation}
	\beta(E, x) = \beta_E \cdot \beta_x.		\nonumber
\end{equation} 
The value of $\beta(E, x)$ depends on the energy deposited in the plastic scintillator during the $\gamma$ photon interaction 
($\beta_E$ factor) and on the position of the $\gamma$ photon interaction along the strip ($\beta_x$ factor). The higher the value of deposited energy, the higher the value of $\beta_E$ parameter and higher the signal amplitude. The $\beta_x$ is necessary to describe absorption of photons propagating through the scintillator strip, since $f_{t_p}$ provides only information about the shape of the signal (see Eq.~(\ref{Eq:f_t_p})). Hence, the closer to the left end of the scintillator, the smaller $x$ (see Fig.~\ref{Fig:exp_setup}) and larger $\beta_x$. Contributions of $\beta_E$ to $\beta$ are the same for both ends of the strip but $\beta_x$ are different. Hereon, in order to simplify the notation of the parameter $\beta(E, x),$ we use only the symbol $\beta.$

\subsection{Reconstruction of the interaction time and position} \label{SubSec:ReconstrTime}

We denote the true values of time and position of $\gamma$ photon interaction with $\Theta^0$ and $x^0,$ respectively, and the corresponding reconstructed values are denoted as $\hat{\Theta}, \hat{x}.$ We add a random noise term $v_{(L,R)}$ to the signal $y_{(L,R)}$ at the left (L) and right (R) end of the strip. Hence a registered signals $\hat{y}_{(L)}$ and $\hat{y}_{(R)}$ may be expressed as:
\begin{align}
	\hat{y}_{(L)}(\Theta^0, x^0) &= y_{(L)}(\Theta^0, x^0) + v_{(L)}. 			\label{Eq:yL_hat} \\
	\hat{y}_{(R)}(\Theta^0, x^0) &= y_{(R)}(\Theta^0, x^0) + v_{(R)}. 			\label{Eq:yR_hat}
\end{align}
We assume that the noise $v_{(L)}$ and $v_{(R)}$ are uncorrelated and obey the same multivariate normal distribution:
\begin{equation}
	v_{(L)}, v_{(R)} \sim \mathcal{N}(0, S), 						\label{Eq:def_v}
\end{equation}
where $S$ is the covariance matrix of $\hat{y}_{(L)}$ and $\hat{y}_{(R)},$  and we introduce notation:
\begin{align}
	\Delta \Theta &= \Theta^0 - \Theta, 			\nonumber			\\
	\Delta x &= x^0 - x.							\nonumber
\end{align} 
According to the definitions of the theoretical ($y$) and registered ($\hat{y}$) signals, the reconstruction of 
$\hat{\Theta}, \hat{x}$ may be pursued by minimization of the function:
\begin{equation}
    W(\Delta\Theta, \Delta x) = (y_{(L)} - \hat{y}_{(L)})(y_{(L)} - \hat{y}_{(L)})^T + (y_{(R)} - \hat{y}_{(R)})(y_{(R)} - \hat{y}_{(R)})^T. 						\label{Eq:f_celu}
\end{equation}  
The solutions $\hat{\Theta}, \hat{x}$ are found as:
\begin{equation}
    (\Delta\hat{\Theta}, \Delta\hat{x}) = \arg \min W(\Delta\Theta, \Delta x) 			\label{Eq:arg_min_f_celu}
\end{equation}  
where hat denotes the estimators.

From Eqs.~(\ref{Eq:yL_hat})-(\ref{Eq:yR_hat}) and~(\ref{Eq:f_celu}) it is seen that the error function $W$ is a positive-valued random variable.
In order to determine $\Delta\hat{\Theta}$ we assume that error of time of interaction has normal distribution: 
\begin{equation}
	\Delta \hat{\Theta} \sim \mathcal{N}(0, \sigma_{\Theta}^2), 			\label{Eq:def_distr_delta_theta}
\end{equation}
where the $\sigma_{\Theta}$ is a searched time resolution of the J-PET instrument.

\subsection{Determination of time resolution} \label{Sec:DetTimeRes}

In order to calculate the time resolution, $W$ has to be examined near the minimum,~$(0, 0).$ According to Eq.~(\ref{Eq:f_celu}), the random variable $W(0, 0)$ may be expressed as: 
\begin{align}
	W(0, 0) &= v_{(L)}v_{(L)}^T + v_{(R)}v_{(R)}^T, \nonumber \\
	&= \sum_{n=1}^N v_{(L)}^2(n) + v_{(R)}^2(n).	\label{Eq:W_00}  					
\end{align}
The variance of $W$ in the minimum will be denoted hereafter as $\text{Var}[W_{\text{min}}].$ Using Eq.~(\ref{Eq:def_v}) and assuming the diagonality of matrix $S,$ yields:
\begin{equation}
 	\text{Var}[W_{\text{min}}]= 2 \sum_{n=1}^N 2 S^2(n,n).      \label{Eq:Var_W_00}
\end{equation}

On the other hand, we may analyse the shape of the function $W$ in the two-dimensional space of time $(\Delta\hat{\Theta})$ and position $(\Delta \hat{x})$ errors. For the purpose of this work, we will consider only the $(\Delta\hat{\Theta})$ error, and therefore analyse $W$ in one dimension $(\Delta \hat{x}=0).$ Taylor series expansion of $W$ around  $(0, 0)$ is given as: 
\begin{align}
	W(\Delta \hat{\Theta}, 0) &\approx W(0,0) +  \frac{\partial W(0, 0)}{\partial \Delta\hat{\Theta}} \Delta\hat{\Theta} + 
	\frac{1}{2} \cdot \frac{\partial^2 W(0, 0)}{\partial \Delta\hat{\Theta}^2} \Delta\hat{\Theta}^2		\nonumber \\
	&\approx \alpha_0 + \alpha_1 \Delta\hat{\Theta} + \alpha_2 \Delta\hat{\Theta}^2.
\end{align}
It is evident that the first two coefficients ($\alpha_0, \alpha_1$) are equal to zero and the quadratic approximation simplifies to:
\begin{equation}
	W(\Delta \hat{\Theta}, 0) \approx \alpha_2 \Delta\hat{\Theta}^2. \label{Eq:W_approx}
\end{equation}
Under the assumption of normality of $\Delta\hat{\Theta}$ distribution, see Eq.~(\ref{Eq:def_distr_delta_theta}), the random variable $W(\Delta \hat{\Theta}, 0)$ given in Eq.~(\ref{Eq:W_approx}), has a $\chi^2$ distribution with the variance:  
\begin{equation}
 	\text{Var}[W_{\text{min}}] \approx 2 {\alpha_2}^2 {\sigma_{\Theta}}^4.      \label{Eq:Var_W_a_theta}
\end{equation}
The comparison of two formulas describing the $\text{Var}[W_{\text{min}}],$ in Eq.~(\ref{Eq:Var_W_a_theta}) and (\ref{Eq:Var_W_00}), enable us to determine time resolution, defined as the standard deviation $\sigma_{\Theta}:$
\begin{equation}
	\sigma_{\Theta} = \sqrt[4]{\frac{2\sum_{n=1}^N S^2(n,n)}{{\alpha_2}^2}}.	\label{Eq:sigma_theta}
\end{equation}

\subsection{Determination of coincidence resolving time} \label{Sec:DetCRT}

In order to facilitate the direct comparison with results published in the field of TOF-PET we will evaluate CRT based on the time resolution ($\sigma_{\Theta}$). In the first approximation CRT equals to $2.35\sqrt{2} \cdot \sigma_{\Theta}.$ However, a fundamental lower limit of the CRT is defined by the time spread due to the unknown depth-of-interaction (DOI) in a single scintillator. It should be stressed that this factor gains importance for large scintillator detectors as in e.g. J-PET. Since the interactions may occur with nearly equal probability along the whole thickness ($d$) of the plastic scintillator, time spread in a single scintillator may be well approximated by the uniform distribution with the width of $d/c,$ where $c$ denotes the 
$\gamma$ photon speed. This implies that the distribution of the time difference between two detected $\gamma$ photons has a triangle form with FWHM equal to $d/c.$ Therefore, the final value of CRT may be estimated with the formula:
\begin{equation}
	\text{CRT} = \sqrt{11.04 \cdot \sigma_{\Theta}^2 + \frac{d^2}{c^2}  }.			\label{Eq:CRT_1}
\end{equation}
As seen from Eqs.~(\ref{Eq:sigma_theta})-(\ref{Eq:CRT_1}), in order to evaluate $\sigma_{\Theta}$ and therefore CRT, one has to know the shape of pdf function $f_{t_r},$ to calculate the $\alpha_2$ coefficient, and also the errors of the signal registered on the photomultipliers, to calculate the covariance matrix $S$. Determination of the shape of $f_{t_r}$ was discussed in the previous section. In the next section we will analyse the sources of errors in the signals $\hat{y}_{(L)}, \hat{y}_{(R)}.$

\subsection{Analysis of registered signals errors} 
\label{Sec:AnalysisErrors}

The noise contribution to the signals registered on the left ($\hat{y}_{(L)}$) and right ($\hat{y}_{(R)}$) side of the scintillator strip is the same, and therefore in this section we will skip the $L, R$ indices. In further analysis we assume that the noise signal $v,$ see Eq.~(\ref{Eq:yL_hat}), is defined as a sum of two components:
\begin{equation}
	v = v_p + v_r,	\label{Eq:error_v}			
\end{equation}
where $v_p$ describes the perturbations of the pdf function $f_{t_r},$ based on limited number of input photon signals, and $v_r$ stands for the signal recovery noise. The latter component is introduced by the procedure of signal recovery based on the limited number of registered samples of the signal in the voltage domain. The problem of signal recovery was widely discussed in Ref. (Raczynski \etal 2015a, b). We assume that the noises $v_p$ and $v_r$ are uncorrelated and normally distributed with covariance matrices $S_p$ and $S_r,$ respectively. Thus, one may write that:
\begin{equation}
	S = S_p + S_r.	\label{Eq:covar_S}			
\end{equation}
The exact values of $v_p$ and $v_r$ depend on the type of the photomultiplier applied. In this work we consider two types of photomultipliers:
\begin{itemize}
\item PMT - vacuum tube photomultiplier treated as a basic one in the current J-PET prototype,
\item MCP  - microchannel plates photomultiplier.
\end{itemize}
It should be underlined that the following analysis does not include the silicon photomultipliers. We have provided the extensive research of the possibilities of the application of the silicon photomultiplier in the J-PET tomograph in our previous study in Ref.~(Moskal \etal 2016).

Noises $v_p$ and $v_r$ are mainly influenced by the width of the single photoelectron contributing to the final output signal, and quantum efficiency of the photomultiplier. The most distinctive feature of the MCP photomultiplier is the capability of the registration of arrival time of each photon. Thus, the output signal may be evaluated by using a model of the single photon. For all types of photomultipliers we use the Gaussian model (Bednarski \etal 2014) for shape of signal of single photoelectron, with the width $\sigma_p.$ In the experimental section we will optimize $\sigma_p$ parameter for the MCP photomultiplier, aiming to minimize the $v_p$ noise. The quantum efficiency may be used directly to estimate the number of photoelectrons induced in the photomultiplier, $N_p.$ In the following we will apply $N_p,$ to model the total output signal. 

It is worth noting that $v_r$ vanishes in the case of MCP photomultiplier. There is no need to recover the output signal since all arrival times of photons are registered. In the following we will shortly describe the noises $v_p$ and $v_r.$

\subsubsection{Analysis of $v_p$.}

The registered signal $y$ affected only by the $v_p$ noise will be denoted with:
\begin{equation}
	\tilde{y} = y + v_p.							\nonumber					
\end{equation}
The signal $\tilde{y}$ consists of $N_p$ signals from individual photoelectrons:
\begin{equation}
	\tilde{y} = \sum_{k=1}^{N_p} \tilde{y_k}. 	\label{Eq:y_tild}
\end{equation}
As mentioned in Section~\ref{Sec:AnalysisErrors}, signal from single photoelectron $\tilde{y_k}$ is assumed to be a Gaussian function:
\begin{equation}
	\tilde{y_k}(n) = \frac{\beta}{\sqrt{(2\pi)}N_p \sigma_p} \exp \left(- \frac{(t^{(n)} - t_r^k)^2}{2\sigma_p^2} \right), 
	\quad \quad n = 1, 2, ..., N, 	\label{Eq:y_k}
\end{equation}
where $t_r^k$ is a random variable with $f_{t_r}$ distribution, that denotes the $k^{th}$ photon's registration time. 

We aim to calculate the diagonal elements of the covariance matrix $S_p:$
\begin{equation}
	S_p(n,n) = E[(\tilde{y}(n) - y(n))^2], \quad \quad n = 1, 2, ..., N,	\label{Eq:element_Sp}			
\end{equation}
where
\begin{align}
	E[(\tilde{y}(n)-y(n))^2] &= E[(\tilde{y}(n) - E[\tilde{y}(n)] +E[\tilde{y}(n)] -y(n))^2]			\nonumber \\
	&=E[(\tilde{y}(n) - E[\tilde{y}(n)])^2]+ (E[\tilde{y}(n)] - y(n))^2								\nonumber \\
	&=\text{Var}(\tilde{y}(n)) + \text{Bias}^2(\tilde{y}(n)), 		\quad \quad n = 1, 2, ..., N.	\label{Eq:final_Var}
\end{align}       
According to the Eq.~(\ref{Eq:y_tild}):
\begin{align}
	E[\tilde{y}(n)] &= N_p \cdot E[\tilde{y_k}(n)], 	 		\label{Eq:Ey} \\
	\text{Var}(\tilde{y}(n)) &= N_p \cdot \text{Var}(\tilde{y_k}(n)),	 &n= 1, 2, ..., N.			\label{Eq:E2y}
\end{align} 
Estimates of the $\text{Var}((\tilde{y}(n))$ and $\text{Bias}(\tilde{y}(n))$ were introduced in Refs. (Rosenblat 1956, Simonoff 1996). Assuming that the underlying pdf function $f_{t_r}$ is sufficiently smooth, and that $\sigma_p \rightarrow 0$ with $N_p \sigma_p \rightarrow \infty$ as $N_p \rightarrow \infty,$ the Taylor series expansion gives:
\begin{align}
	\text{Bias}(\tilde{y}(n)) &\approx \beta\frac{\sigma_p^2 f_{tr}''(t^{(n)})}{2}, 		 \label{Eq:Bias_lit} \\
	\text{Var}(\tilde{y}(n)) &\approx \beta^2\frac{f_{tr}(t^{(n)})}{2 \sqrt{\pi} N_p \sigma_p}, &n= 1, 2, ..., N, \label{Eq:Var_lit}
\end{align}    
where $f_{tr}''(t^{(n)})$ is a second derivative of the pdf function $f_{tr}(t^{(n)}).$ Above approximations may be inaccurate for finite $N_p.$ The number of registered photoelectrons $N_p$ is of the order of hundreds, and the detailed discussion is given in Sec.~\ref{Sec:ExpSetup}. Therefore, a new method to evaluate the $\text{Var}((\tilde{y}(n))$ and $\text{Bias}(\tilde{y}(n))$ for finite $N_p$ should be proposed. During this study the novel concept of the estimation of a requested statistics has been developed. The method has been described in great details in the Appendix and it was shown that the values of 
$\text{Var}(\tilde{y}), \text{Bias}(\tilde{y})$ may be estimated as:
\begin{align}
	\text{Bias}(\tilde{y}(n)) &\approx \beta
	\left( \frac{2\Phi(t^{(n)}, \lambda\sigma_p)}{3\sqrt{2\pi} \sigma_p}-f_{tr}(t^{(n)}) \right),		\label{Eq:Ey_unif} \\
	\text{Var}(\tilde{y}(n)) &\approx \beta^2\frac{9\Phi(t^{(n)}, \lambda\sigma_p) + 8\Phi^2(t^{(n)}, \lambda\sigma_p) 
	- 16\Phi^3(t^{(n)}, \lambda\sigma_p)}{36 \pi N_p \sigma_p^2}, 	&n= 1, 2, ..., N,		\label{Eq:E2y_unif}
\end{align}    
where $\lambda$ is the parameter defining the range of the second argument of function $\Phi$:
\begin{equation}
	\Phi(t^{(n)}, \lambda\sigma_p) = F_{t_r}(t^{(n)} + \lambda\sigma_p) - F_{t_r}(t^{(n)} - \lambda\sigma_p), \quad \quad
	 n= 1, 2, ..., N,
	\label{Eq:Phi_t}
\end{equation}
and $F_{t_r}(t^{(n)})$ is the cumulative distribution function of $f_{t_r}(t^{(n)})$ calculated at $t^{(n)}.$ 
Discussion of formulas~(\ref{Eq:Ey_unif}, \ref{Eq:E2y_unif}) is given in the Appendix.

It should be underlined that both estimation methods, proposed (Eqs.~(\ref{Eq:Ey_unif}, \ref{Eq:E2y_unif})) and based on Taylor series approximation (Eqs.~(\ref{Eq:Bias_lit}, \ref{Eq:Var_lit})), have the same asymptotic properties. It may be shown that for $\sigma_p \rightarrow 0$ with $N_p \sigma_p \rightarrow \infty$ as $N_p \rightarrow \infty:$
\begin{align}
	\text{Bias}(\tilde{y}(n)) &= 0, 		 \nonumber \\
	\text{Var}(\tilde{y}(n)) &= 0,  \quad n= 1, 2, ..., N. \nonumber
\end{align}    

\subsubsection{Analysis of $v_r$.} \label{Sec:AnalysisVr}

Denote the signal $y$ affected only by the $v_r$ noise as:
\begin{equation}
	\hat{y} = \tilde{y} + v_r.	\label{Eq:wr_simple}											
\end{equation}
The recovery process takes place only provided the complete output $\tilde{y}$ is registered on a photomutliplier. If times of photon's arrival are registered, as in the MCP photomultiplier, the $v_r = 0.$ Recovery of the signal $\hat{y}$ is carried out only for the PMT photomultiplier. 
 
The details of signal recovery process were given in Ref. (Raczynski \etal 2015a), and here only the main points will be recalled. The evaluation of the signal $\hat{y}$ requires two steps: (i) recovery of the sparse expansion $\hat{x}$ and (ii) calculation of $\hat{y}$ based on the $\hat{x}$. The relation between the solution $\hat{y}$ and its sparse representation 
$\hat{x}$ is linear:
\begin{equation}
	\hat{y} = A \hat{x},	\label{Eq:estY}
\end{equation}
where $A$ is an orthonormal matrix. As it was shown in Ref. (Raczynski \etal 2015a), from the Bayes theory the properties of regularized solution can be found, in particular its covariance matrix, denoted hereafter as $S_{r(x)}$, may be easily derived:
\begin{equation}
	S_{r(x)} = \left(P^{-1} + \frac{M}{\sigma^2 N} \textbf{1} \right)^{-1}
	\label{eq:Sx}
\end{equation}
where $P$ is the covariance matrix of the sparse signals $x$, and $M$ denotes the number of registered samples of the signals 
$y, \sigma$ is the standard deviation of the measurement error. Finally, based on Eq.~(\ref{Eq:estY}), the covariance matrix 
$S_r$ is given:
\begin{equation}
	S_r = A \left(P^{-1} + \frac{M}{\sigma^2 N} \textbf{1} \right)^{-1} A^T. 	\label{Eq:Sr}
\end{equation}
 
\section{Experimental results}

\subsection{Experimental setup} \label{Sec:ExpSetup}

\begin{figure}[h!]
	\centerline{\includegraphics[width=0.7\textwidth]{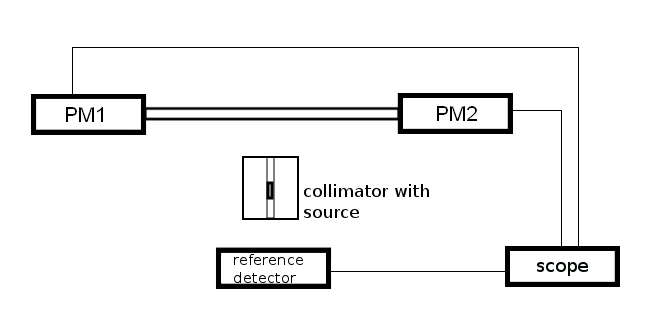}}
\caption{
Scheme of the experimental setup. 
\label{Exper_setup}
}
\end{figure}

In this section we investigate the accuracy of the proposed method for evaluation of the time resolution and CRT. The model is validated by performing the experiment with a single detection module of the J-PET scanner built out from the BC-420 plastic scintillator strip, with dimensions of 5 x 19 x 300~mm, read out at two ends by the Hamamatsu R4998 (PMT) photomultipliers. 
Our experimental setup is depicted in Fig.~\ref{Exper_setup}. 
Measurements are performed using $\gamma$ photon from the $^{22}$Na source placed inside the lead collimator between the scintillator strip and the reference detector. The reference detector consists of a small scintillator strip with a thickness of 4~mm. Collimated beam emerging through 1.5~mm wide and 20~cm long slit is used for irradiating desired points across the strip. In order to detect the event, a coincident registration of signals from the PM1 and a reference detector is required. Such trigger conditions enable us to select precisely the annihilation quanta reducing the background from the deexcitation photon (1.27 MeV) to the negligible level (Moskal \etal 2014b). The time of triggering by the reference detector is used to estimate the event arrival time. The constant electronic time delay between the true event time and measured arrival time to the reference detector does not influence the time resolution and is shifted to zero. The full waveforms of PMT signals are sampled using the Lecroy SDA 6000A oscilloscope running at a 20 GSps sampling rate. 

In our previous studies it was shown that the time resolution is fairly independent of the irradiation position (Moskal \etal 2014b). Therefore, we determine the time resolution and CRT of the J-PET scanner in one position, at the center of the strip ($x =$ 15~cm). In order to evaluate the experimental value of time resolution and CRT a data set of $10^4$ pairs of signals from PM1 and PM2 registered in coincidence was analyzed. In the first step, for each pair of fully sampled signals from the left and right ends of the strip, $\tilde{y}_{(L)}$ and 
$\tilde{y}_{(R)},$ a front-end electronic device probing signals at four voltage levels, both at the rising and falling slope, was simulated. The signals $\hat{y}_{(L)}$ and $\hat{y}_{(R)}$ were recovered using 8 samples of signals $\tilde{y}_{(L)}$ and $\tilde{y}_{(R)}$ registered by an oscilloscope, according to the method descried in Sec.~\ref{Sec:AnalysisVr}. 
For the $k$-th pair of the recovered signals $\hat{y}_{(L)}$ and $\hat{y}_{(R)},$ the energy of an event may be estimated based on arithmetic mean of a number of photoelectrons registered at the left and right sides of the scintillator (Moskal \etal 2014b) and is proportional to the sum of integrals of recovered signals $\hat{y}_{(L)}$ and $\hat{y}_{(R)}.$ On the other hand, for the $k$-th pair of $\hat{y}_{(L)}$ and $\hat{y}_{(R)},$ the reconstruction of time ($\hat{\Theta}_k$) and position ($\hat{x}_k$) was pursued by minimization of the function $W$ in Eq.~(\ref{Eq:f_celu}). The value of $\sigma_{\Theta}$ was calculated as the standard deviation of the empirical distribution of $\hat{\Theta}_k$ and was equal to about 80~ps. The corresponding value of CRT calculated based on Eq.~(\ref{Eq:CRT_1}), for a scintillator strip with a thickness of 19~mm, was equal to 275~ps. This value of CRT will be treated as the reference for the proposed approach. For clarity of the presentation, we will calculate in Sec.~3 only the CRT parameter.



According to the scheme presented in Sec.~\ref{Sec:AnalysisErrors}, the evaluation of the time resolution and CRT of the PET system requires the investigation of the parameter $\alpha_2$ and covariance matrix $S$ (see Eq.~(\ref{Eq:sigma_theta})). The values of these parameters vary for a different type of applied photomultipliers and are also sensitive to the position of the point of $\gamma$ photon interaction along the scintillator strip. The values of the parameters will be provided in Sec.~\ref{Sec:ExpSetup} and \ref{Sec:VeriEstMeth}.      


In order to model the signal at the photomultiplier's output, the parameters of three pdf functions $f_{t_e}, f_{t_p}$ and 
$f_{t_d},$ defined in Eq.~(\ref{Eq:f_t_e}), (\ref{Eq:f_t_p}) and (\ref{Eq:f_t_d}), respectively, must be known. It is worth noting that only the last pdf function, $f_{t_d},$ describes the unique properties of a given type of the photomultiplier. The $\sigma_d$ is delivered by the photomultiplier's producer, for Hamamatsu R4998 photomultiplier (PMT) $\sigma_d =$~68~ps and for MCP photomultiplier $\sigma_d =$~40~ps (Hamamatsu~2016). However, as our initial tests show, there is a negligible influence of the $\sigma_d$ value on the performance of the reconstruction method of $\gamma$ photon interaction moment.

\begin{figure}[h!]
	\centerline{\includegraphics[width=0.6\textwidth]{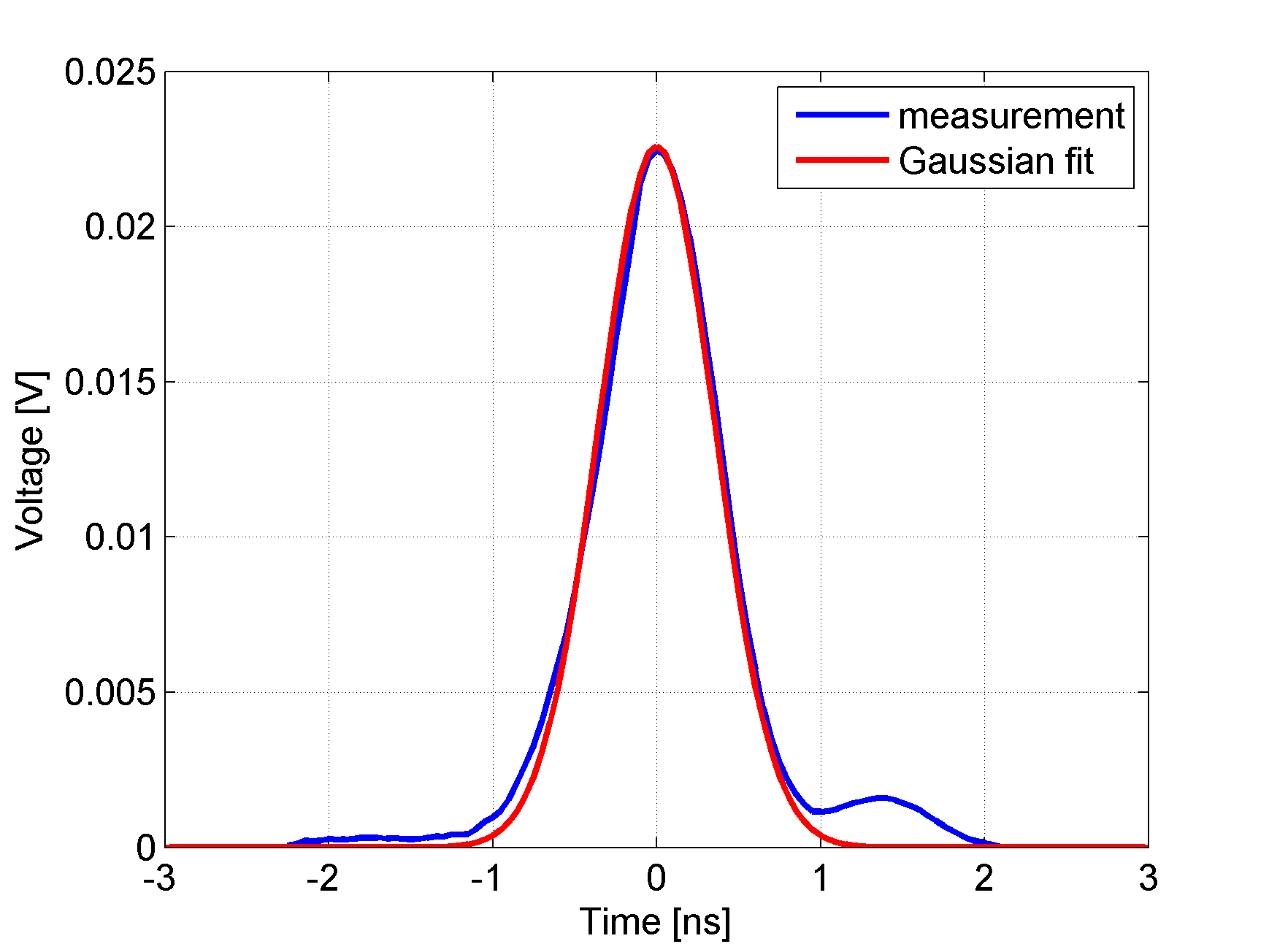}}
\caption{An example of the signal of single photoelectron acquired with PMT photomultiplier (blue curve) and its Gaussian fit (red curve). In the measured signal the two Gaussian are observed, however the second one is much smaller and its influence on the calculated parameters is negligible.  
\label{Fig:single_photon}
}
\end{figure}

The measurement provides a discrete signal $y$ at the photomultiplier. Repeating the measurement of its time under the same condition yields a set of the acquisition times of photoelectrons, cf. Eq.~(\ref{Eq:y_tild}). The signal $y$ consists of $N_p$ Gaussian shaped signals of single photoelectrons. In Fig.~\ref{Fig:single_photon} an example of the single photoelectron signal registered with PMT photomultiplier and its Gaussian fit are shown. The signals are marked with blue and red curves, respectively. The standard deviation $\sigma_p$ of this function is reported in Ref.~(Bednarski \etal 2014) to be equal to 300~ps in the case of PMT photomultiplier. 
However, a different number of photoelectrons ($N_p$) are registered due to the different quantum efficiencies. In the following we will shortly recall the main results of our earlier works enabling us to estimate properly the number $N_p.$ Light yield of plastic scintillators amounts to about 10000 photons per 1 MeV of deposited energy. The 511~keV $\gamma$ photon may deposit maximally 341~keV via Compton scattering (Szymanski \etal 2014), which corresponds to the emission of about 3410 photons. On the other hand in order to decrease the noise due to the scattering of $\gamma$ photon inside patient's body, the minimum energy deposition of about 200 keV is required (Moskal \etal 2012). Therefore, the range of the number of emitted photons discussed hereafter in this article amounts 2000 to 3410. The experiments conducted with PMT photomultipliers revealed that about 280 photoelectrons are produced from the emission of 3410 photons (Moskal \etal 2014). According to the preselected range, from 2000 to 3410 photons, the average number of emitted photons is about 2700. This number corresponds to $N_p =$~220 registered photoelectrons equipped with the PMT photomultiplier. Since the CRT of the J-PET system will be determined at the center of the strip, the numbers of photoelectrons $N_p$ contributing to the signals induced on the left and right scintillator ends are the same and are equal to 110.       

\begin{figure}[h!]
	\centerline{\includegraphics[width=0.6\textwidth]{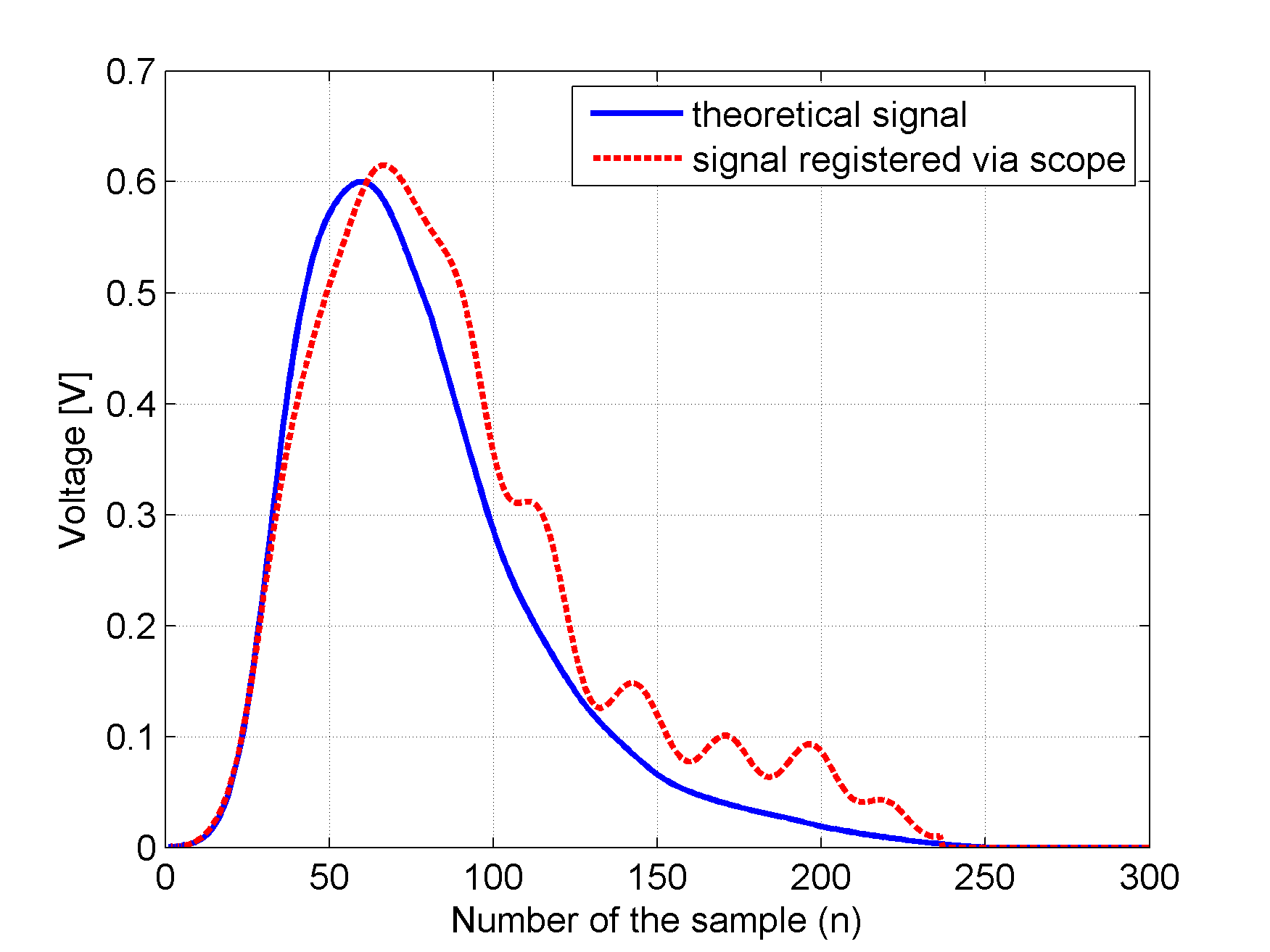}}
\caption{Signals observed on the PMT photomultiplier output generated by interaction in the center of the scintillator strip; theoretical signal $y$ (see Eq.~(\ref{Eq:def_y})) is marked with the blue curve, and an example of signal $\tilde{y}$ registered via oscilloscope (see Eq.~(\ref{Eq:y_tild})) is marked with the red dashed curve (meaning of variable $n$ is the same as in formula~(\ref{Eq:t_n_def})). 
\label{Fig:model_y}
}
\end{figure}

As mentioned at the beginning of Sec.~\ref{Sec:MatMeth}, the values of $\tau_d, \tau_r, \sigma_e$ of the $f_{t_e}$ pdf function were adjusted based on the experimental studies with a single BC-420 scintillator strip. We have provided numerous tests for various strips of the BC-420 scintillator type and we found that the values of the estimated parameters of the $f_{t_e}$ pdf function were consistent within the measurement errors. Therefore, the signals evaluated based on the proposed model, presented in Sec.~\ref{Sec:DetTimeRes}, have shapes very similar to those registered during the experiment via oscilloscope (see Fig.~\ref{Fig:model_y}). In Fig.~\ref{Fig:model_y}, the theoretical signal $y$ at the center of the strip, evaluated from Eq.~(\ref{Eq:def_y}), is presented. The parameter $\beta$ (see formula (\ref{Eq:def_y})) was selected in such way that the amplitude of the signal is equal to the mean amplitude of signals registered at the center of the strip ($x =$ 15~cm). The analytical solution for $f_{t_r}$ function is difficult to find due to the internal convolution in $f_{t_e}$ function (see Eq.~(\ref{Eq:f_t_e})).  Therefore, the numerical evaluation of a convolution operation was applied. The signals $y$ and $\tilde{y}$ in Fig.~\ref{Fig:model_y} are shown in the discrete domain for discrete time samples and the curves connecting points are plotted to guide ones eye. 

\begin{figure}[h!]
	\centerline{\includegraphics[width=0.6\textwidth]{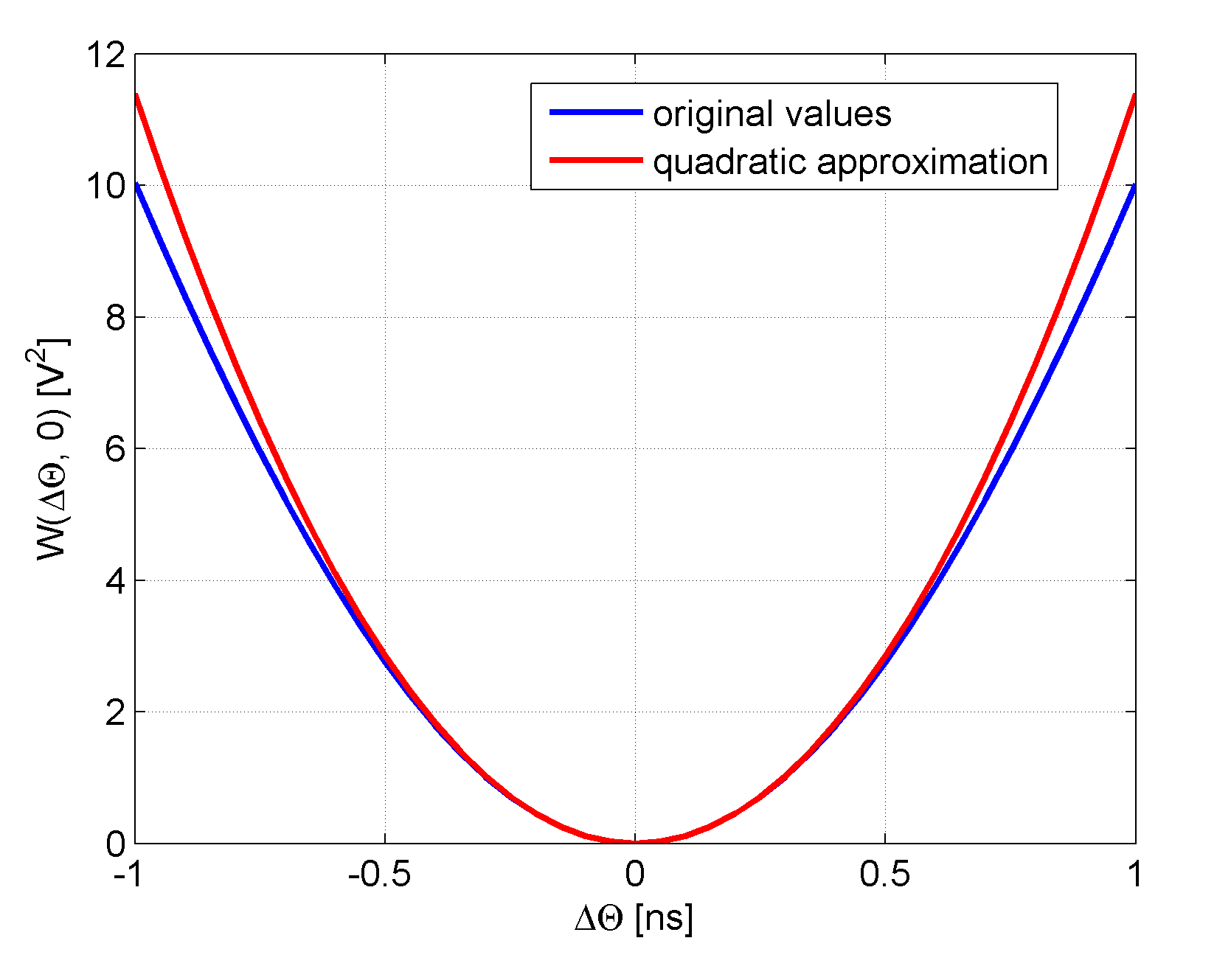}}
\caption{The shape of $W(\Delta \Theta, 0)$ near to the minimum.
\label{Fig:W_experiment}
}
\end{figure}

The information about the signal $y$ may be directly applied to evaluate the value of the parameter $\alpha_2.$ In this work we are interested only in determination of CRT of the J-PET system and we assume that the position of the $\gamma$ photon interaction is known exactly (see Eq.~(\ref{Eq:W_approx})). Therefore, for a fixed position of the interaction, the signal $y$ may be shifted only in time domain due to the error of time measurement $\Delta \Theta.$ For $\Delta \Theta=$~0 the theoretical and registered signals overlap and $W(\Delta \Theta, 0)=$~0, see Eq.~(\ref{Eq:f_celu}). In order to evaluate $\alpha_2,$ the error $\Delta \Theta$ was varied from -1 to 1~ns. For each value of $\Delta \Theta,$  function $W(\Delta \Theta, 0)$ was evaluated based on the shape of signal $y$ shown in Fig.~\ref{Fig:model_y}. The resulting, experimental function $W(\Delta \Theta, 0)$ is presented in Fig.~\ref{Fig:W_experiment} with blue curve (see also Eq.~(\ref{Eq:f_celu})).

According to Eq.~(\ref{Eq:W_approx}), the experimental function $W(\Delta \Theta, 0)$ may be approximated near 
$\Delta \Theta=$~0 with the quadratic function. The quadratic approximation of the 
$W(\Delta \Theta, 0)$ function is marked in Fig.~\ref{Fig:W_experiment} with the red curve and the coefficient of the second order polynomial function is equal to 11.2~$\frac{\text{V}^2}{\text{ns}^2}.$

\subsection{Verification of signal $\tilde{y}$ estimation method} \label{Sec:VeriEstMeth}

According to the assumptions in Sec.~\ref{Sec:AnalysisErrors}, two main contributors to the signal's noise are $v_p$ and $v_r.$ The $v_r$ was estimated in Ref. (Raczynski \etal 2015a) and will be recalled at the end of this point. In this section, a detailed study of the approximation method of the $\text{Var}(\tilde{y}(n))$ and $\text{Bias}(\tilde{y}(n))$ will be carried out. The proposed method, see Eq.~(\ref{Eq:Ey_unif}) and (\ref{Eq:E2y_unif}), will be compared with the well known approximation technique based on the Taylor series expansion, see Eq.~(\ref{Eq:Bias_lit}) and (\ref{Eq:Var_lit}). As the reference for the results of both analytical approaches, the Monte-Carlo (MC) simulation will be provided.        

\begin{figure}[h!]
	\centerline{\includegraphics[width=0.6\textwidth]{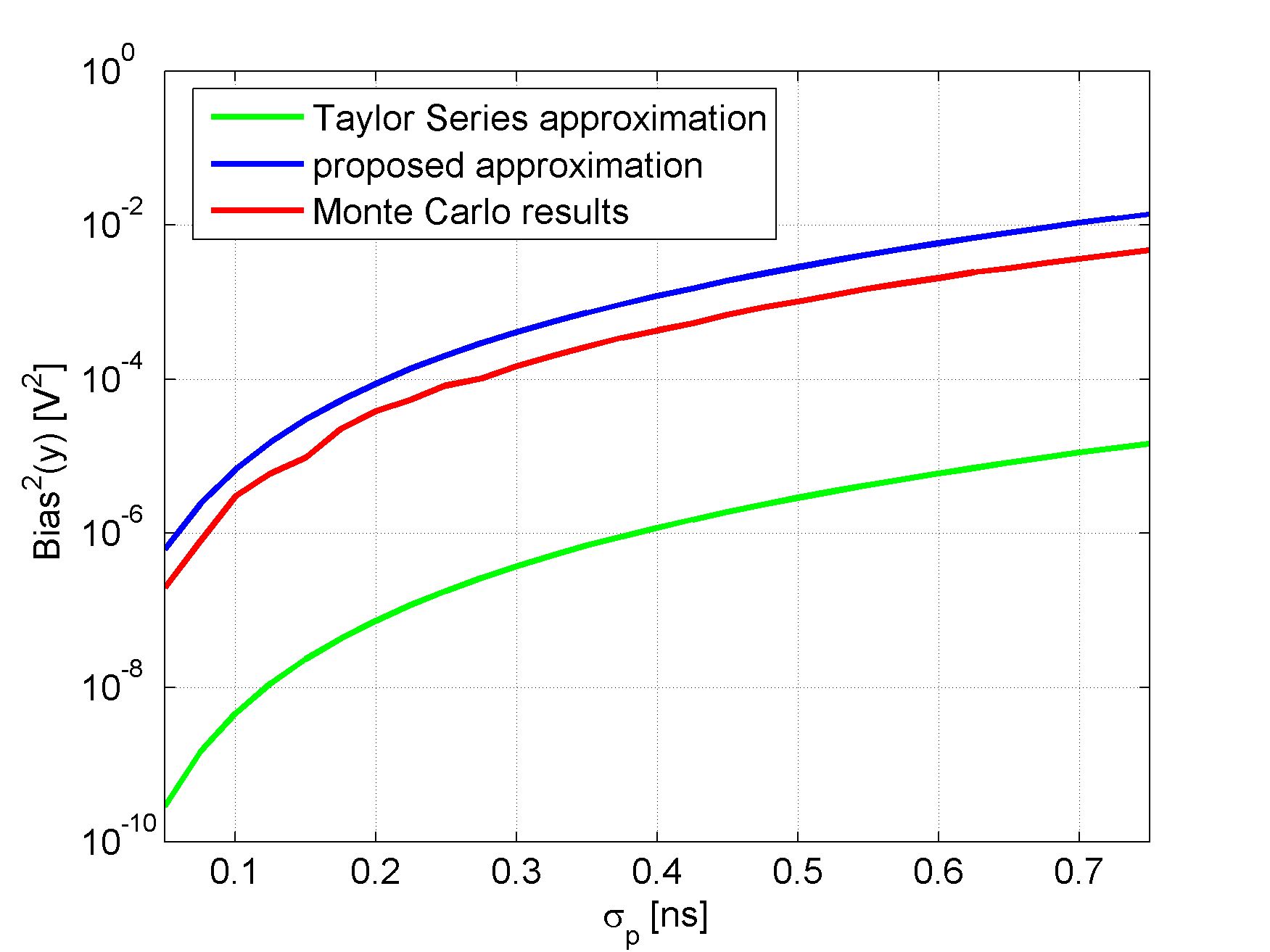}}
\caption{The comparison of estimation of $\text{Bias}^2[\tilde{y}]$ with two analytical approaches: the proposed one (blue curve), and based on the Taylor series expansion (green curve). The reference characteristics was obtained with the Monte-Carlo simulation (red curve).  
\label{Fig:comparison_bias}
}
\end{figure}

\begin{figure}[h!]
	\centerline{\includegraphics[width=0.6\textwidth]{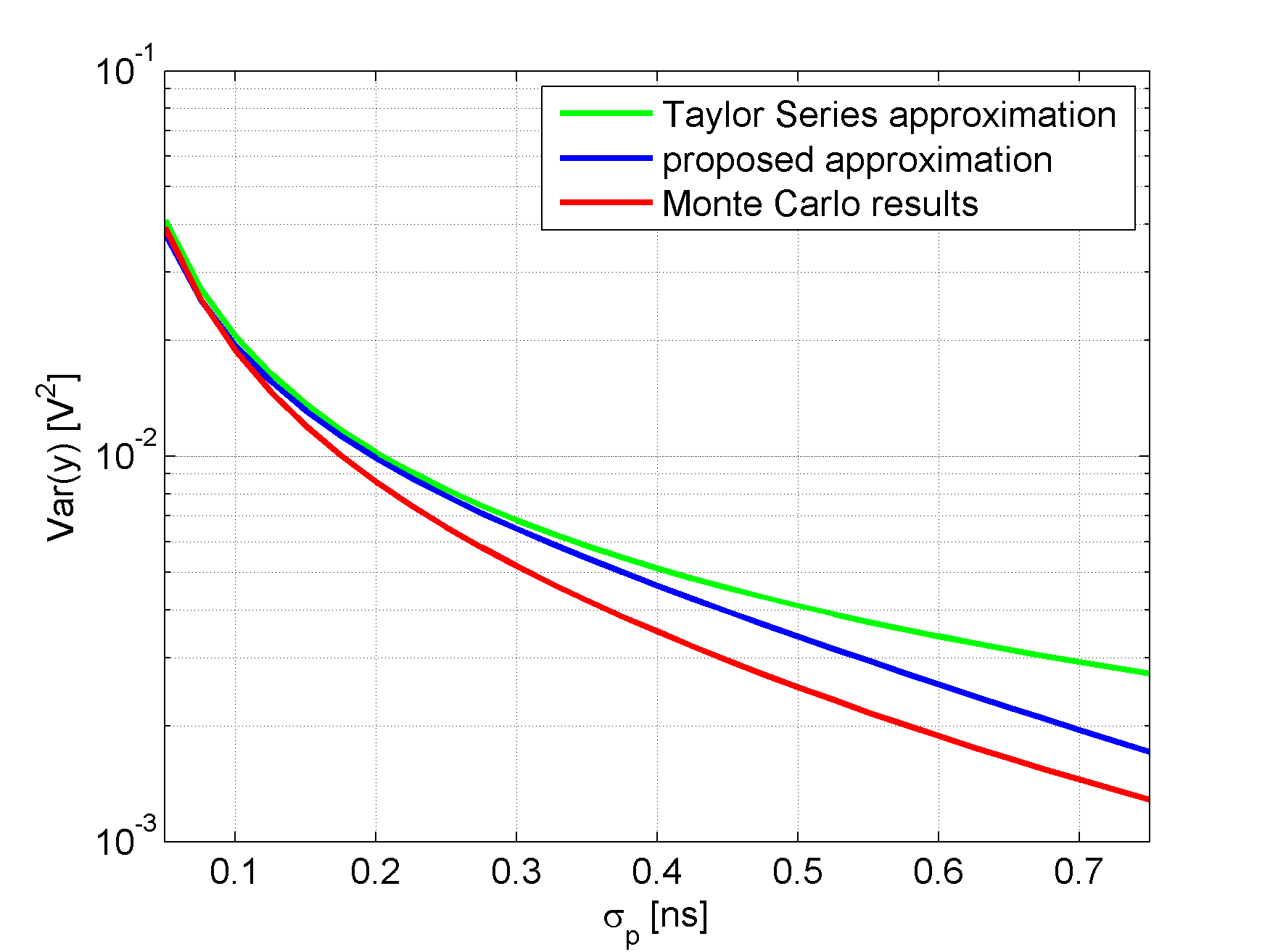}}
\caption{The comparison of estimation of $\text{Var}[\tilde{y}]$ with two analytical approaches: the proposed one (blue curve), and based on the Taylor series expansion (green curve). The reference characteristics was obtained with the Monte-Carlo simulation (red curve).  
\label{Fig:comparison_var}
}
\end{figure}

The MC simulation was carried out for the constant number of photoelectrons $N_p =$~220, registered by the PMT photomultiplier. In order to simulate the $\text{Var}(\tilde{y}(n))$ and $\text{Bias}(\tilde{y}(n)),$ only one timestamp of the original signal $y,$ corresponding to the maximum value of 0.6~V (see Fig.~\ref{Fig:model_y}), was used. 
The analysis of the maximum value in signal $y$ allows one to evaluate the main contribution in the covariance matrix $S_p;$ as seen from 
Fig.~\ref{Fig:comparison_diag_Sp_Sr},  the location of maximum value of the signal $y$ corresponds to the location of maximum value on diagonal of the covariance matrix $S_p$.
The maximum value of the original signal $y$ is observed in the sample $n =$~60 (see Fig.~\ref{Fig:model_y}). In the first step of MC simulation the random values of photons registration times $t_r^k$ ($k = 1, 2, ..., N_p$) were selected according to the $f_{t_r}$ distribution. Next, the values of all $N_p$ functions $\tilde{y_k}(60)$ were evaluated based on the Eq.~(\ref{Eq:y_k}) and summed up giving $\tilde{y}(60).$  The above-mentioned procedure was repeated $10^6$ times for different values of $\sigma_p$ from 50~ps to 750~ps with step 25~ps. 
The range of $\sigma_p$ has been selected after a preliminary calculations taking into account the expected number of registered photoelectrons in the J-PET scenario. 
Based on the large number of samples of $\tilde{y}(60),$ the accurate estimation of bias and variance was possible. The resulting $\text{Bias}^2(\tilde{y}(60))$ and $\text{Var}((\tilde{y}(60))$ are shown in Fig.~\ref{Fig:comparison_bias} and \ref{Fig:comparison_var}, respectively.


The reference values of $\text{Bias}^2(\tilde{y}(60))$ and $\text{Var}(\tilde{y}(60)),$ obtained with MC simulation, are marked with red curves in Fig.~\ref{Fig:comparison_bias} and \ref{Fig:comparison_var}, respectively. An approximation of 
$\text{Var}(\tilde{y})$ for the proposed method and method based on the Taylor series expansion (blue and green curves, respectively) are very similar to reference curve for small values and tend to differ for larger values of $\sigma_p.$ However, in the most interesting region, for $\sigma_p$ equal to about 300~ps, the proposed method is more accurate than the Taylor series based method and the values of $\text{Var}(\tilde{y})$ are equal to 6.5$\times 10^{-3}$~V$^2$ and 7.0$\times 10^{-3}$~V$^2,$ respectively (the reference value of $\text{Var}(\tilde{y})$ from MC simulation is equal to 5.1$\times 10^{-3}$~V$^2$). Comparison of the $\text{Bias}^2(\tilde{y})$ and $\text{Var}(\tilde{y})$ curves reveals the fundamental relation between variance and bias. The variance dominates for smaller values of $\sigma_p$ and becomes comparable with bias for $\sigma_p$ at the level of about 500~ps (compare two reference, red curves in Fig.~\ref{Fig:comparison_bias} and \ref{Fig:comparison_var}). For $\sigma_p$ larger than 500~ps, the total error is mostly influenced by the bias. It is worth noting that in that case the Taylor series based method significantly underestimates the values of $\text{Bias}^2(\tilde{y}),$ see Fig.~\ref{Fig:comparison_bias}, which leads to the underestimation of the overall error.   

\subsection{Evaluation of time resolution of the J-PET system} \label{Sec:EvaluaTimeRes}

\begin{figure}[h!]
	\centerline{\includegraphics[width=0.6\textwidth]{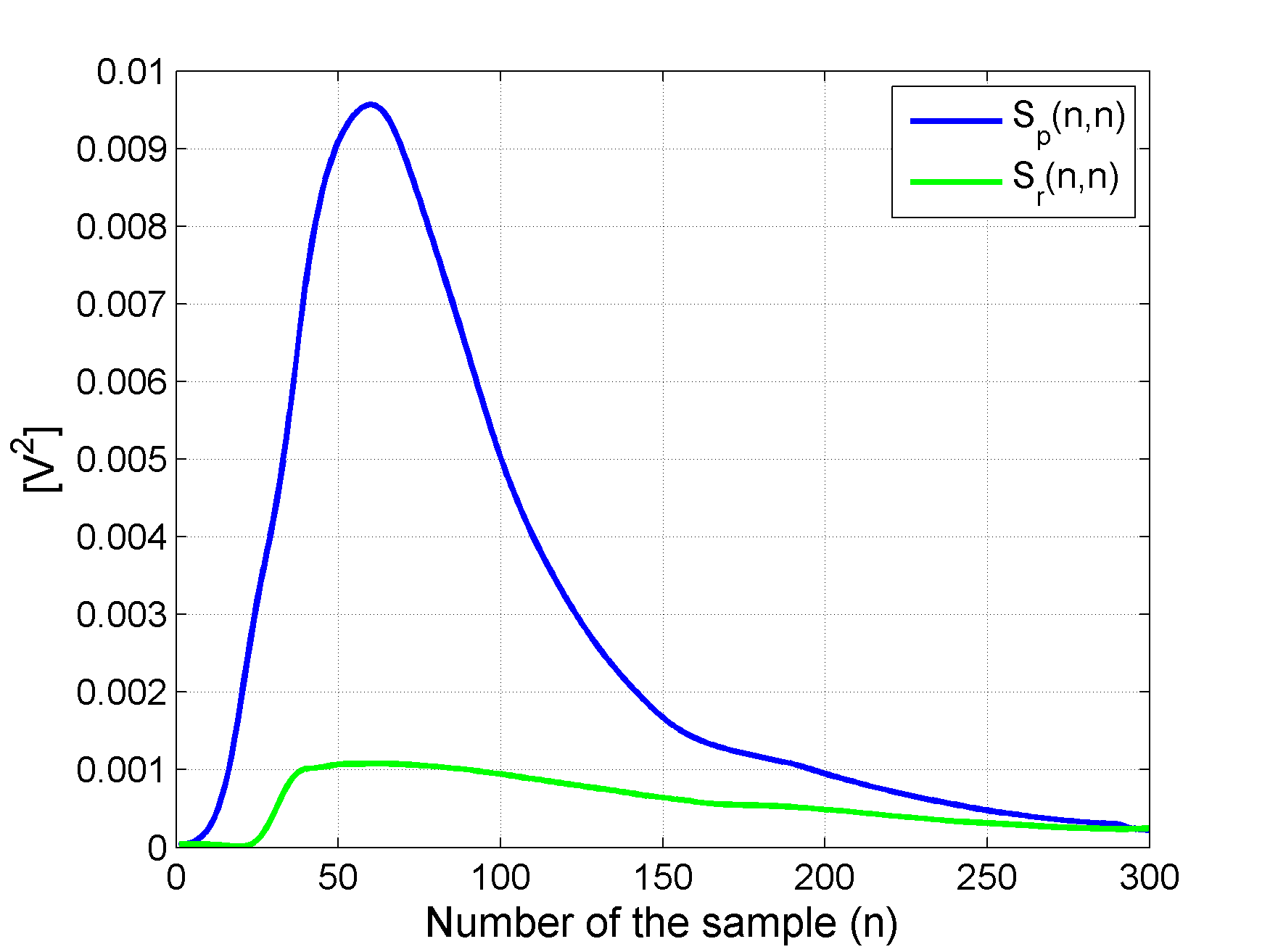}}
\caption{Comparison of the diagonal elements of the covariance matrices $S_p$ and $S_r$ for the PMT photomultiplier.  
\label{Fig:comparison_diag_Sp_Sr}
}
\end{figure}

In the first step we compare the covariance matrices $S_p$ and $S_r$ (see Eqs.~(\ref{Eq:error_v}) and (\ref{Eq:covar_S})) according to the description in Sec.~\ref{Sec:ExpSetup} and using our previous study (Raczynski \etal 2015a). The resulting values of the diagonal elements of $S_p$ and $S_r$ are shown in Fig.~\ref{Fig:comparison_diag_Sp_Sr}. 
Theoretical values of $S_p$ were evaluated as in Sec.~\ref{Sec:VeriEstMeth} for PMT and the results are marked in 
Fig.~\ref{Fig:comparison_diag_Sp_Sr} with blue curve. The values of $S_r$ elements are marked with green curve in Fig.~\ref{Fig:comparison_diag_Sp_Sr}. The comparison of the resulting characteristics with the shape of the pdf function $f_{t_r},$ presented in Fig.~\ref{Fig:model_y}, indicates that the reconstructed errors $S_p$ and $S_r$ are highly related to the signal value. The maximal values of the diagonal elements of $S_p$ and $S_r$ occur near to the maximum of signal $y$ 
(Fig.~\ref{Fig:model_y}). On the other hand, the analysis of the characteristics plotted in Fig.~\ref{Fig:comparison_diag_Sp_Sr} reveals that the error introduced by the limited number of photoelectrons in the registered signal ($S_p$) is a dominating factor. 

\begin{figure}[h!]
	\centerline{\includegraphics[width=0.6\textwidth]{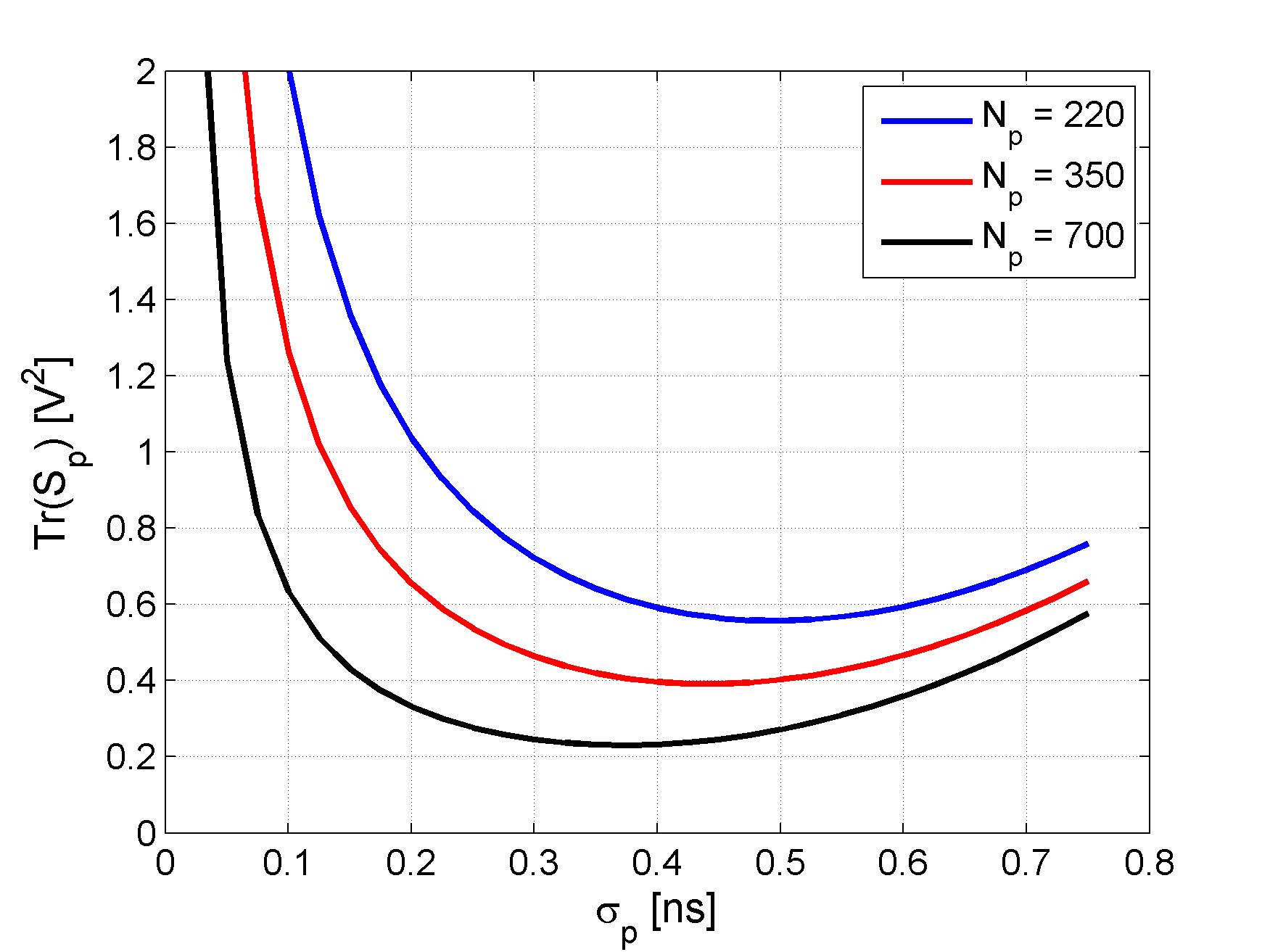}}
\caption{Trace of the $S_p$ matrix as a function of the standard deviation of single photoelectron signal ($\sigma_p$) for three specified numbers of registered photoelectrons 
$N_p = $~220, 350 and 700.
\label{Fig:Trace_Sp}
}
\end{figure}

In order to compare the reconstructed values of the covariance matrices $S_p$ and $S_r,$ we use the trace ($\textnormal{Tr}$)
of the covariance matrix since the diagonality is assumed. The values of $\textnormal{Tr}(S_p)$ and $\textnormal{Tr}(S_r)$ for different photomultipliers type are gathered in Tab.~\ref{Table:Summ}. In the following we will analyse the value of $\textnormal{Tr}(S_p)$ as the function of the number of registered photoelectrons ($N_p$) and standard deviation of the single photoelectron signal ($\sigma_p$). The resulting characteristics of $\textnormal{Tr}(S_p)$ as a function of $\sigma_p$ are shown in Fig.~\ref{Fig:Trace_Sp}. 
The values of $\textnormal{Tr}(S_p)$ were calculated for three specified numbers of registered photoelectrons 220, 350 and 700 and are marked in Fig.~\ref{Fig:Trace_Sp} with blue, red and black curves, respectively. The smallest value of $N_p$ is specific for PMT photomultiplier, as mentioned in Sec.~\ref{Sec:ExpSetup}. The highest number $N_p =$~700 indicates the maximal number of registered photoelectrons in the experimental scenario, and was selected in order to demonstrate the best theoretical resolution of the J-PET. Results in Fig.~\ref{Fig:Trace_Sp} show that all the $\textnormal{Tr}(S_p)$ functions, evaluated for a given number of registered photoelectrons, have a minimum. The shape of the $\textnormal{Tr}(S_p)$ functions illustrates the fundamental trade-off between variance and bias, as mentioned in Sec.~\ref{Sec:VeriEstMeth}. Hence, for given number of registered photoelectrons, it is possible to adjust the optimal value of $\sigma_p$, denoted hereafter with $\sigma_{p(opt)}.$ Comparison of the $\sigma_{p(opt)}$ values, for three $N_p$ numbers in Fig.~\ref{Fig:Trace_Sp}, shows that the larger the number of registered photoelectrons the smaller the value of $\sigma_{p(opt)}.$ For instance, for the PMT photomultiplier that registers 220 photoelectrons on average, the minimum error occurs for the $\sigma_{p(opt)} = $~500~ps (blue curve in Fig.~\ref{Fig:Trace_Sp}). In the case of PMT photomultiplier the $\sigma_{p}$ is not a variable and has fixed value of about 300~ps. However, the MCP photomultiplier registers timestamps of the signal instead of the complete signal. Therefore, the value of $\sigma_{p(opt)}$ of each contributing signal may be adjusted accordingly to the number of registered timestamps ($N_p$). In that sense, the optimization of the $\sigma_{p(opt)}$ value for MCP photomultiplier may be provided. Simulations using $f_{t_r}$ function provide $N_p^{-4.1}$ dependence of $\sigma_{p(opt)}$ on the number of photoelectrons.

\begin{figure}[h!]
	\centerline{\includegraphics[width=0.6\textwidth]{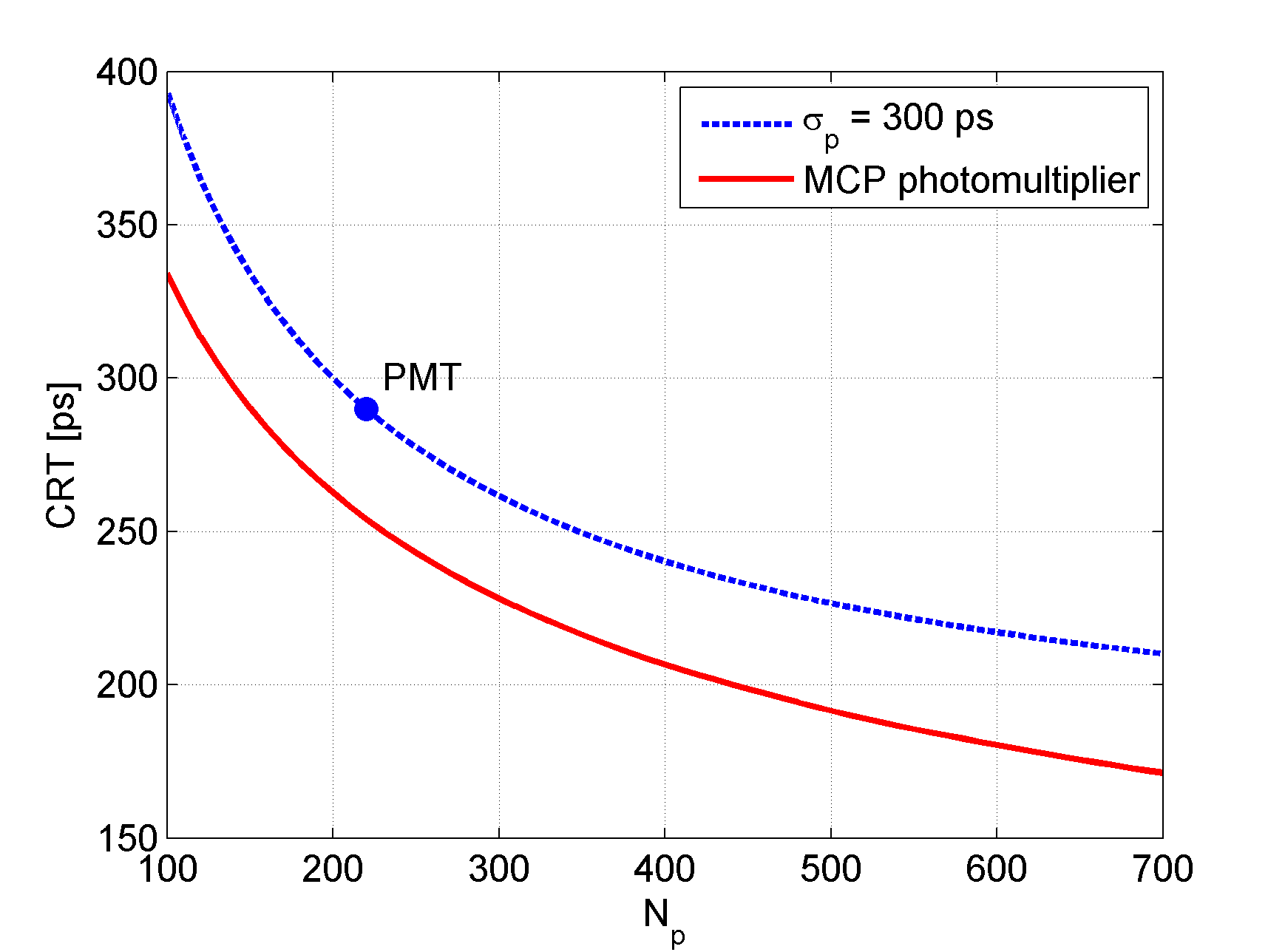}}
\caption{Theoretical calculations of CRT versus the number of photoelectrons $N_p$, of the J-PET tomograph equipped with two types of photomultipliers: PMT and MCP.  The presented
values of CRT take into account an additional smearing of the time due to the unknown depth of interaction in a scintillator strip with a thickness of 19~mm, see Eq.~(\ref{Eq:CRT_1}) for details. 
\label{Fig:sigma_theta}
}
\end{figure}

In general, the MCP photomultiplier is capable to register all the timestamps of the photons reaching the scintillator end. In order to account for possible inefficiency of the MCP, we determine the characteristics of the J-PET equipped with the MCP in the range $100 \leqslant N_p \leqslant 700.$ First, for a given number $N_p,$ the optimal value of $\sigma_{p(opt)}$ was estimated based on the characteristics of $\textnormal{Tr}(S_p)$ (see Fig.~\ref{Fig:Trace_Sp}). Next, the matrix $S_p$ was calculated based on the proposed technique, see Eq.~(\ref{Eq:Ey_unif}) and (\ref{Eq:E2y_unif}). Finally, $\sigma_{\Theta}$ was  evaluated based on Eq.~(\ref{Eq:sigma_theta}). In the case of MCP, $\textnormal{Tr}(S_r) = $~0, since the output signal is given directly based on the measured timestamps and assumed shape of the single photoelectron signal. Resulting characteristics of CRT is given with red solid line in Fig.~\ref{Fig:sigma_theta}. The presented values of CRT take into account an additional smearing of the time due to the unknown depth of interaction in a scintillator strip with a thickness of 19~mm, see Eq.~(\ref{Eq:CRT_1}) for details. 

Additionally, the CRT calculated for $\sigma_p = $~300~ps (a shape of single photoelectron signal characteristic for PMT photomultiplier), including also the $v_r$ error introduced by the signal recovery procedure, $\textnormal{Tr}(S_r) = $~0.22~V$^2$ reported in Ref.~(Raczynski \etal 2015a), is marked with blue dashed line in 
Fig.~\ref{Fig:sigma_theta}. The theoretical value of CRT for the PMT photomultiplier is marked with full circle on the blue curve, for $N_p = $~220. The theoretical CRT of the J-PET scanner with the PMT photomultiplier is about 290~ps and agrees with the experimental value of CRT, reported to be about 275~ps (Raczynski \etal 2015a). For fixed value of the quantum efficiency (equivalent to the number $N_p$), further improvement of CRT is possible by the application of the MCP photomultipliers. In the wide range of numbers of registered photoelectrons shown in Fig.~\ref{Fig:sigma_theta}, improvement of about 30~ps is observed (red and blue dashed curves). The presented results show that the best theoretical CRT of the J-PET scanner, with 30~cm long strips, estimated for the MCP photomultiplier that is capable to register all timestamps of arrival for 700 photons, is at the level of 170~ps. The main results of the study as well as the parameters of the analysed photomultipliers, are summarized in Tab.~\ref{Table:Summ}.

\begin{table}[ht]
\caption{Summary of theoretical CRT calculations of the J-PET scanner. The parameters are described in the text.  The presented values of CRT take into account an additional smearing of the time due to the unknown depth of interaction in a scintillator strip with a thickness of 19~mm, see Eq.~(\ref{Eq:CRT_1}) for details. 
} 
\centering 
\begin{tabular}{l | l | c | ccc} 
\hline\hline 
Parameter & Unit & \multicolumn{4}{c}{Photomultiplier type} \\ [0.5ex] 
\hline 
 &  & PMT  &\multicolumn{3}{c}{MCP} \\ [0.5ex] 
\hline 
$N_p$ & 1 & 220 & 220 & 350 & 700 \\ 
$\sigma_p$ & ps & 300 & 500 & 420 & 360  \\
$\sigma_{p(opt)}$ & ps & 500 & 500 & 420 & 360\\
Tr$(S_p)$ & V$^2$ & 0.73 & 0.55 & 0.39 & 0.23 \\
Tr$(S_r)$ & V$^2$ & 0.22 & 0.0 & 0.0 & 0.0 \\ 
Tr$(S)$ & V$^2$ & 0.95 & 0.55 & 0.39 & 0.23  \\ 
CRT & ps & $\textbf{290}$ & $\textbf{260}$ & $\textbf{215}$ & $\textbf{170}$ \\ [1ex] 
\hline 
\end{tabular}
\label{Table:Summ} 
\end{table}

\section{Extension of the proposed method for conventional PET systems}


The proposed framework for calculation of the time resolution and CRT may also be applied for the state of the art PET scanners equipped with crystal scintillators. To this purpose, in order to estimate the time resolution, function $W,$ introduced in Eq.~(\ref{Eq:f_celu}), has to be adopted to the new situation. First of all, during the reconstruction of the time of 
$\gamma$ photon interaction ($\hat{\Theta}$), only signals from one side of the crystals are acquired; in J-PET scanner two signals are registered at both ends of scintillator. Moreover, due to the small size of the crystals, only reconstruction of the interaction time $\hat{\Theta}$ is being carried out; function $W$ is one-dimensional. Therefore, the function $W$ is defined as:
\begin{equation}
    W(\Delta\Theta) = (y- \hat{y})(y - \hat{y})^T  			\nonumber			
\end{equation}  
and finally, the standard deviation $\sigma_{\Theta}$ is given with the formula:
\begin{equation}
	\sigma_{\Theta} = \sqrt[4]{\frac{\sum_{n=1}^N S^2(n,n)}{{\alpha_2}^2}},		\label{Eq:sigma_theta_CRYST}
\end{equation}
which differs from the Eq.~(\ref{Eq:sigma_theta}) only with a factor $\sqrt[4]{2}.$ In order to evaluate time resolution and CRT, one has to evaluate the distribution time of the photon registration at the photomultiplier ($t_r$), to calculate the $\alpha_2$ coefficient, and also the errors of the signal registered on the photomultipliers, to calculate the covariance matrix $S.$


In case of  PET scanners with inorganic crystal scintillators, the description of the photon registration time $t_r$ at the photomultiplier includes two random components $t_e$ and $t_d:$
\begin{equation}
	t_r = t_e + t_d. 	\nonumber
\end{equation}
In comparison to the J-PET, here the propagation time of the photon along the scintillator ($t_p$) may be neglected due to small size of the single crystal.
Therefore, the only difference in calculation of $t_r$ is the evaluation of distribution of $t_e.$ The main parameter that governs the speed of light emission after the absorption of a 
$\gamma$ photon is the decay time. Crystal materials show decay pulse shapes that are single- or multi-exponential. For example, for the BGO crystals the bi-exponential shape of the distribution of time $t_e$ is observed (Seifert \etal 2012).
 
On the other hand, determination of time resolution, defined in Eq.~(\ref{Eq:sigma_theta_CRYST}), requires the information about the covariance matrix $S$ of the registered signal 
$\hat{y}.$ In Sec.~\ref{Sec:AnalysisErrors} we derived the analytical description of the main components of the covariance matrix $S,$ i.e. matrices $S_p$ and $S_r.$ 

The formula for calculation of elements of $S_p$ matrix, describing the perturbations of the distribution function $f_{t_r}$ based on a limited number of registered photoelectrons, given in Eq.~(\ref{Eq:element_Sp}), may be also applied to the PET scanners with crystal scintillators. The only differences are in the parameters describing the shape of the $f_{t_r}$ distribution function and the expected number of photoelectrons ($N_p$) while including the light yield of crystal scintillators.
        
The latter component, matrix $S_r,$ is introduced by the procedure of signal recovery based on the limited number of registered samples of the signal in the voltage domain. 
The J-PET system involves a four-threshold sampling method to generate samples of a signal waveform. An example of a similar electronic system for probing the signals in a voltage domain, coupled with experimental setup equipped with LSO crystals was developed in Ref.~(Kim \etal 2009). The waveforms of signals were read-out by the oscilloscope and the electronic system for probing these signals in a voltage domain with four thresholds was applied to reconstruct the pulse shape. This scheme allows to evaluate all the parameters  required to calculate the signal recovery error, according to the formula given in Eq.~(\ref{Eq:Sr}), for the PET system with crystal scintillators. 

\section{Conclusions}


In this paper, we introduced a new method of estimation of time resolution and CRT of the J-PET system using only simulations which were tested based on the data from a single detector module. This is particularly useful for design of expensive device. In case of  J-PET tomograph the most expensive part of the system are the photomultipliers. In this work two types of photomultipliers were simulated: the vacuum tube photomultipliers and microchannel plates.   
 

The basic idea of the method is the use of the statistical nature of the whole signal acquisition process. We have highlighted three statistical phenomena: the emission of photons in the scintillator strip, the propagation of light pulses along the strip and registration of light in photomultipliers. Parameters of the probability density functions were selected in order to describe properly light pulses from the plastic scintillator BC-420.    


An important aspect of our work concerns the statistical analysis of an error of reconstruction of the probability density function based on the set of single photoelectrons signals. In this work dependences of an overall variance and bias on the number and width of the single photoelectron signals were evaluated. The proposed estimation method was validated by using the Monte Carlo simulation and it was shown that obtained results are consistent. Moreover, the proposed technique was demonstrated to be more accurate than the approach from the literature (Rosenblat 1956, Simonoff 1996). The developed estimation scheme is general and may be incorporated elsewhere.


In the experimental section, the method of time resolution and CRT estimation was tested using signals registered by means of the single detection module of the J-PET scanner. In order to evaluate a CRT of the J-PET detector, we have incorporated the method described in Ref. (Raczynski \etal 2015a). In the cited work, the CRT obtained with the experimental scheme with vacuum tube photomultipliers, was reported to be equal to about 275~ps. Our calculation shows that the application of the proposed estimation method can give very similar result of about 290~ps. The consistency of the experimental and theoretical results, obtained for the J-PET scanner equipped with the vacuum tube photomultipliers suggests that the estimated CRTs for other photomultipliers are reliable. The determined CRTs for the detector with microchannel plates amount to 215~ps and 170~ps assuming the 50$\%$ and 
100$\%$ quantum efficiency of photomultiplier, respectively.  


Future work will address investigation of other aspects of signal acquisition process by using the proposed statistical model, e.g. the influence of the parameters of distribution of the photon emission time on time resolution. In this study, the parameters of the distribution were selected in order to describe the properties of light signals observed in the BC-420 plastic scintillator. However, our group develops a novel type of plastic scintillator and examines the influence of the chemical composition of the plastic scintillator on the overall performance of the J-PET detector (Wieczorek \etal 2015a, 2015b, 2016). Application of the proposed model to that task, enables us to use information about the shape of the distribution of the time of photon emission directly to predict the CRT of the J-PET detector.

\ack

We acknowledge technical and administrative support of A. Heczko, M. Kajetanowicz, W. Migda\l{} and the financial support by the Polish National Center for Development and Research through grant INNOTECH-K1/IN1/64/159174/NCBR/12, the Foundation for Polish Science through MPD programme, the EU and MSHE Grant No. POIG.02.03.00-161 00-013/09, Doctus - the Lesser Poland PhD Scholarship Fund, and Marian Smoluchowski Krak\'ow Research Consortium "Matter-Energy-Future".  

\appendix

\section{Kernel density estimation}

The function $\tilde{y_k},$ describing the $k^{th}$ signal from a single photoelectron, given in  Eq.~(\ref{Eq:y_k}), may be approximated with:
\begin{equation}
 \tilde{y_k}(n) \approx
  \begin{cases}
    \frac{\beta}{\sqrt{(2\pi)}N_p \sigma_p} \left(1 - \frac{(t^{(n)} - t_r^k)^2}{\lambda^2 \sigma_p^2} \right) & \quad 
    t_r^k \in (t^{(n)} - \lambda\sigma_p, t^{(n)} + \lambda\sigma_p)\\
    0  & \quad \text{otherwise} \\
  \end{cases}
  \label{Eq:y_k_approx}
\end{equation}
where $n = 1, 2, ..., N,$ and $\lambda$ contributes to the signal width. The probability that the random variable $\tilde{y_k}(n)$ is equal to the specified value, may be calculated based on the previously introduced function $\Phi$, see Eq.~(\ref{Eq:Phi_t}). In particular, the probability that the random variable $\tilde{y_k}(n) = 0$ is equal to 
$1 - \Phi(t^{(n)}, \lambda\sigma_p);$ the $k^{th}$ registration time $t_r^k$ is out of range  
$(t^{(n)} - \lambda\sigma_p, t^{(n)} + \lambda\sigma_p),$ see the second case in Eq.~(\ref{Eq:y_k_approx}). Denoting the first case in Eq.~(\ref{Eq:y_k_approx}) with $u_k:$  
\begin{equation}
	u_k(n) =\frac{\beta}{\sqrt{(2\pi)}N_p \sigma_p} \left(1 - \frac{(t^{(n)} - t_r^k)^2}{\lambda^2 \sigma_p^2} \right), 
	\quad n = 1, 2, ..., N,
\end{equation}
we may write that for $n = 1, 2, ..., N,$ the expected value of $\tilde{y_k}(n)$ is equal:
\begin{align}
	E[\tilde{y_k}(n)] &= E[u_k(n)]\Phi(t^{(n)}, \lambda\sigma_p) + E[0](1-\Phi(t^{(n)}, \lambda\sigma_p))	\nonumber \\
	 &= E[u_k(n)]\Phi(t^{(n)}, \lambda\sigma_p),			\label{Eq:dod:Eyk}
\end{align} 
and the variance of $\tilde{y_k}(n)$ is equal:
\begin{align}
	\text{Var}(\tilde{y_k}(n)) &= E[(u_k(n) - E[u_k(n)])^2]\Phi(t^{(n)}, \lambda\sigma_p) + 
          E[(0 - E[u_k(n)])^2](1-\Phi(t^{(n)}, \lambda\sigma_p))			\nonumber \\
	 &= \text{Var}(\tilde{u_k}(n))\Phi(t^{(n)}, \lambda\sigma_p) + 
	 E[u_k(n)]^2(1-\Phi(t^{(n)}, \lambda\sigma_p)).			\label{Eq:dod:Varyk}
\end{align} 
In order to simplify the further calculations the following assumption is proposed. Note that in most interesting cases the range $(t^{(n)} - \lambda\sigma_p, t^{(n)} + \lambda\sigma_p),$ is narrow in comparison to the estimated pdf function $f_{t_r}$ domain. Therefore, the pdf function $f_{t_r}$ is considered to be uniform in the range $(t^{(n)} - \lambda\sigma_p, t^{(n)} + \lambda\sigma_p):$ 
\begin{equation}
	f_{t_r}(\epsilon) \simeq \text{const.}  \quad \epsilon 
	\in (t^{(n)} - \lambda\sigma_p, t^{(n)} + \lambda\sigma_p).
	\label{Eq:unif}
\end{equation}
It is worth noting that the smaller is the ratio of the single- to overall signal width, the better is the performance of the proposed approximation method. 

Under the assumption in Eq.~(\ref{Eq:unif}), required moments in Eqs.~(\ref{Eq:dod:Eyk}) and (\ref{Eq:dod:Varyk}), 
$E[u_k(n)],$ $E[u_k(n)]^2$ and $\text{Var}(\tilde{u_k}(n))$, can be easily derived. After some simple calculations the equations for the expected value and the variance of the random variable $\tilde{y_k}(n)$ are given by formulas:
\begin{align}
	E(\tilde{y}(n)) &\approx \beta\frac{2\Phi(t^{(n)}, \lambda\sigma_p)}{3\sqrt{2\pi} \sigma_p},			
	&n= 1, 2, ..., N,		\nonumber \\
	\text{Var}(\tilde{y}(n)) &\approx \beta^2\frac{9\Phi(t^{(n)}, \lambda\sigma_p) + 8\Phi^2(t^{(n)}, \lambda\sigma_p) 
	- 16\Phi^3(t^{(n)}, \lambda\sigma_p)}{36 \pi N_p \sigma_p^2}, 	&n= 1, 2, ..., N.		\nonumber
\end{align}    

\section*{References} 

\begin{harvard}

\item[] Bednarski T \etal 2014 Calibration of photomultipliers gain used in the J-PET detector {\it Bio-Algorithms and Med-Systems} {\bf 10} 13 [arXiv:1312.2744 [physics.ins-det]]

\item[] Candes E, Romberg J, Tao T 2006 Robust uncertainty principles: Exact signal reconstruction from highly incomplete frequency information {\it IEEE Transactions on Information Theory} {\bf 52} 489 [ arXiv:math/0409186v1 [math.NA]]

\item[] Conti M 2009 State of the art and challenges of time-of-flight PET {\it Physics in Medicine} {\bf 25} 1

\item[] Conti M 2011 Focus on time-of-flight PET: the benefits of improved time resolution {\it Eur. J. Nucl. Med. Mol. Imaging} {\bf 38} 1147


\item[] DeGroot M 1986 Probability and Statistics {\it Reading, MA: Addison-Wesley} 420

\item[] Donoho D 2006 Compressed sensing {\it IEEE Transactions on Information Theory} {\bf 52} 1289

\item[] Eljen Technology 2016 http://www.eljentechnology.com

\item[] Hamamatsu 2016 http://www.hamamatsu.com

\item[] Humm J L, Rosenfeld A and Del Guerra A 2003 From PET detectors to PET scanners  {\it Eur. J. Nucl. Med.
Mol. Imaging} {\bf 30} 1574

\item[] Karp J \etal 2008 Benefit of time-of-flight in PET: experimental and clinical results {\it J. Nucl. Med.}
 {\bf 49} 462

\item[] Kim H \etal 2009 A multi-threshold sampling method for TOF-PET signal processing {\it Nucl. Instrum. and Methods in Phys. Res. A} {\bf 602} 618

\item[] Moskal P \etal 2011 Novel detector systems for the Positron Emission Tomography 
  {\it Bio-Algorithms and Med-Systems} {\bf 7} 73 [arXiv:1305.5187 [physics.med-ph]]

\item[] Moskal P \etal 2012 TOF-PET detector concept based on organic scintillators
  {\it Nuclear Medicine Review} {\bf 15} C81 [arXiv:1305.5559 [physics.ins-det]]

\item[] Moskal P \etal 2014a A novel TOF-PET detector based on organic scintillators
 {\it Radiotheraphy and Oncology Procedings Supplement} {\bf 110} S69

\item[] Moskal P \etal 2014b Test of a single module of the J-PET scanner based on plastic scintillators
 {\it Nucl. Instrum. and Methods in Phys. Res. A} {\bf 764} 317 [arXiv:1407.7395 [physics.ins-det]]

\item[] Moskal P \etal 2015 A novel method for the line-of-response and time-of-flight reconstruction in
TOF-PET detectors based on a library of synchronized model signals {\it Nucl. Instrum. and Methods in Phys. Res. A} {\bf 775} 54 [arXiv:1412.6963 [physics.ins-det]]

\item[] Moskal P \etal 2016 Time resolution of the plastic scintillator strips with matrix photomultiplier readout
for J-PET tomograph {\it Physics in Medicine and Biology} {\bf 61} 2025 [arXiv:1602.02058 [physics.ins-det]]

\item[] Moszynski M and Bengtson B 1977 Light pulse shapes from plastic scintillators {\it Nucl. Instrum. and Methods in Phys. Res. A} {\bf 142} 417 

\item[] Moszynski M and Bengtson B 1979 Status of timing with plastic scintillation detectors {\it Nucl. Instrum. and Methods in Phys. Res. A} {\bf 158} 1 

\item[] Palka M \etal 2014 A novel method based solely on FPGA units enabling measurement of time and
charge of analog signals in positron emission tomography {\it Bio-Algorithms and Med-Systems} {\bf 10} 41 [arXiv:1311.6127 [physics.ins-det]]

\item[] Parzen E 1962 On estimation of a probability density function and mode {\it Annals of Mathematical Statistics} {\bf 33} 1065

\item[] Raczynski L \etal 2014 Novel method for hit-position reconstruction using voltage signals in plastic
scintillators and its application to positron emission tomography {\it Nucl. Instrum. and Methods in Phys. Res. A} {\bf 764} 186 [arXiv:1407.8293 [physics.ins-det]]

\item[] Raczynski L \etal 2015a Compressive sensing of signals generated in plastic scintillators in a novel J-PET
instrument {\it Nucl. Instrum. and Methods in Phys. Res. A} {\bf 786} 105 [arXiv:1503.05188 [physics.ins-det]]

\item[] Raczynski L \etal 2015b Reconstruction of Signal in Plastic Scintillator of PET using Tikhonov
Regularization {\it Proc. IEEE Engineering in Medicine and Biology Conference} 2784

\item[] Rosenblat M 1956 Remarks on some nonparametric estimates of a density function {\it Annals of Mathematical Statistics} {\bf 27} 832

\item[] Saint Gobain Crystals 2016 http://www.crystals.saint-gobain.com

\item[] Seifert S, van Dam H and Schaart D 2012 The lower bound on the timing resolution of scintillation
detectors {\it Physics in Medicine and Biology} {\bf 57} 1797

\item[] Simonoff J 1996 Smoothing methods in statistics {\it Springer Verlag} New York

\item[] S\l{}omka P, Pan T, Germano G 2016 Recent advances and future progress in PET instrumentation {\it Semin. Nucl. Med.} 
{\bf 46} 5

\item[] Spanoudaki V and Levin C 2011 Investigating the temporal resolution limits of scintillation detection from pixellated elements: comparison between experiment and simulations {\it Physics in Medicine and Biology} {\bf 56} 735

\item[] Szymanski K \etal 2014 Simulations of gamma quanta scattering in a single module of the J-PET
detector {\it Bio-Algorithms and Med-Systems} {\bf 10} 71 [arXiv:1312.0250 [physics.ins-det]]

\item[] Tikhonov A 1963 Soviet mathematics {\it Doklady} {\bf 4} 1035

\item[] Tikhonov A and Arsenin V 1977 Solutions of Ill-Posed Problems {\it Winston and Sons, Washington, D.C.}

\item[] Townsend D 2004 Physical principles and technology of clinical PET imaging {\it Ann. Acad. Med. Singap.} {\bf 33} 133

\item[] Wieczorek A \etal 2015a PALS investigations of free volumes thermal expansion of J-PET plastic scintillator synthesized in polystyrene matrix {\it NUKLEONIKA} {\bf 60} 777 [arXiv:1508.06820 [physics.ins-det]]

\item[] Wieczorek A \etal 2015b A pilot study of the novel J-PET plastic scintillator with 2-(4-styrylphenyl)benzoxazole as a wavelength shifter {\it Acta Phys. Pol A} {\bf 127} 1487 [arXiv:1502.02901 [physics.ins-det]]

\item[] Wieczorek A \etal 2016 Novel plastic scinitllators for the fully digital and MRI compatible J-PET scanner {\it Physica Medica: European Journal of Medical Physics} {\bf 32} 232

\end{harvard}


\end{document}